\documentclass[fleqn,usenatbib]{mn2e}
\usepackage{aas_macros}

\usepackage{eso-pic}

\AddToShipoutPictureBG*{%
  \AtPageUpperLeft{%
    \hspace{0.75\textwidth}%
    \raisebox{-3.5\baselineskip}{%
      \makebox[0pt][l]{\textnormal{DES-2020-0536}}
}}}%

\AddToShipoutPictureBG*{%
  \AtPageUpperLeft{%
    \hspace{0.75\textwidth}%
    \raisebox{-4.5\baselineskip}{%
      \makebox[0pt][l]{\textnormal{FERMILAB-PUB-21-448-AE}}
}}}%

\usepackage{todonotes}

\usepackage[T1]{fontenc}
\usepackage{ae,aecompl}

\usepackage{graphicx}	
\usepackage{amsmath}	
\usepackage{amssymb}	
\usepackage{xspace}
\usepackage{lineno}
\usepackage{layouts}
\usepackage[normalem]{ulem}  
\usepackage{newtxtext,newtxmath}
\usepackage{hyperref}
\usepackage[noabbrev]{cleveref}

\newcommand{\code}[1]{\texttt{#1}}
\newcommand{\balrog}{\code{Balrog}}

\newcommand{\D}{\mathrm{d}}
\newcommand{\Hor}{\mathrm{H}}
\newcommand{\hyperrank}{\textsc{hyperrank}\xspace}
\newcommand{\buzzard}{\textsc{Buzzard}\xspace}
\newcommand{\sompz}{\textsc{SOMPZ}\xspace}
\newcommand{\deltaz}{\ensuremath{\Delta z}\xspace}
\newcommand{\nz}{\ensuremath{n(z)}\xspace}
\newcommand{\nzfid}{\ensuremath{n_{\tt{Fid}}(z)}\xspace}
\newcommand{\sigmadeltaz}{\ensuremath{\sigma(\Delta z)}\xspace}
\newcommand{\multinest}{\textsc{MultiNest}\xspace}
\newcommand{\polychord}{\textsc{PolyChord}\xspace}
\newcommand{\emcee}{\textsc{emcee}\xspace}
\newcommand{\cosmosis}{\textsc{CosmoSIS}\xspace}

\newcommand{\photoz}{photo-\emph{z}\xspace}

\definecolor{jpcolor}{rgb}{0.0, 0.5, 0.0}

\newcommand{\response}[1]{#1}

\title[DES-Y3: Marginalisation over redshift uncertainties]{Dark Energy Survey Year 3 results: Marginalisation over redshift distribution uncertainties using ranking of discrete realisations}

\author[J. P. Cordero et al.]{
\parbox{\textwidth}{
\Large{Juan P. Cordero$^{1}$\thanks{E-mail: juan.cordero@postgrad.manchester.ac.uk},
Ian Harrison$^{2,1}$\thanks{E-mail: ian.harrison-2@manchester.ac.uk},
Richard P. Rollins$^{3,1}$,
G. M. Bernstein$^{4}$,
S. L. Bridle$^{1}$,
A. Alarcon$^{5,6,7}$,
O. Alves$^{8,9}$,
A. Amon$^{10,11,12}$,
F. Andrade-Oliveira$^{9,13}$,
H. Camacho$^{9,13}$, 
A. Campos$^{14}$, 
A. Choi$^{15}$, 
J. DeRose$^{16,17}$, 
S. Dodelson$^{14}$, 
K. Eckert$^{4}$, 
T. F. Eifler$^{18,19}$, 
S. Everett$^{16}$, 
X. Fang$^{19}$, 
O. Friedrich$^{20,21}$, 
D. Gruen$^{10,11,12}$, 
R. A. Gruendl$^{22,23}$, 
W. G. Hartley$^{24}$, 
E. M. Huff$^{18}$, 
E. Krause$^{19}$, 
N. Kuropatkin$^{25}$, 
N. MacCrann$^{26}$, 
J. McCullough$^{11}$, 
J. Myles$^{10,11,12}$, 
S. Pandey$^{4}$, 
M. Raveri$^{27}$, 
R. Rosenfeld$^{13,28}$, 
E. S. Rykoff$^{11,12}$, 
C. S\'anchez$^{4}$, 
J. S\'anchez$^{25}$,
I. Sevilla-Noarbe$^{29}$, 
E. Sheldon$^{30}$, 
M. Troxel$^{31}$, 
R. Wechsler$^{10,11,12}$, 
B. Yanny$^{25}$, 
B. Yin$^{14}$,
Y. Zhang$^{25}$,
M. Aguena$^{13,32}$, 
S. Allam$^{25}$, 
E. Bertin$^{33,34}$, 
D. Brooks$^{35}$, 
D. L. Burke$^{11,12}$, 
A. Carnero Rosell$^{36,37}$, 
M. Carrasco Kind$^{22,23}$, 
J. Carretero$^{38}$, 
F. J. Castander$^{6,7}$, 
R. Cawthon$^{39}$, 
M. Costanzi$^{40,41}$, 
L. da Costa$^{13,32}$, 
M. E. da Silva Pereira$^{8}$,
J. De Vicente$^{29}$, 
H. T. Diehl$^{25}$,
J. Dietrich$^{42}$, 
P. Doel$^{35}$, 
J. Elvin-Poole$^{15,43}$, 
I. Ferrero$^{44}$, 
B. Flaugher$^{25}$, 
P. Fosalba$^{6,7}$, 
J. Frieman$^{25,27}$, 
J. Garcia-Bellido$^{45}$, 
D. Gerdes$^{8,46}$,
J. Gschwend$^{13,32}$, 
G. Gutierrez$^{25}$, 
S. Hinton$^{47}$, 
D. L. Hollowood$^{16}$, 
K. Honscheid$^{15,43}$, 
B. Hoyle$^{42,48,49}$, 
D. James$^{50}$, 
K. Kuehn$^{51,52}$, 
O. Lahav$^{35}$, 
M. A. G. Maia$^{13,32}$, 
M. March$^{4}$,
F. Menanteau$^{22,23}$, 
R. Miquel$^{38,53}$, 
R. Morgan$^{39}$, 
J. Muir$^{11,54}$, 
A. Palmese$^{25,27}$, 
F. Paz-Chinchon$^{23,55}$, 
A. Pieres$^{13,56}$, 
A. Plazas Malag\'on$^{57}$, 
E. S\'anchez$^{29}$, 
V. Scarpine$^{25}$, 
S. Serrano$^{6,7}$, 
M. Smith$^{58}$, 
M. Soares-Santos$^{8}$, 
E. Suchyta$^{59}$, 
M. Swanson$^{23}$,
G. Tarle$^{8}$,
D. Thomas$^{60}$,
C. To$^{10,11,12}$, 
and
T. N. Varga$^{48,49}$ (DES Collaboration)
}
\parbox{\textwidth}{ \small
\textit{The authors' affiliations are shown at the end of this paper.}}}}

\date{Accepted XXX. Received YYY; in original form ZZZ}

\pubyear{2021}

\hypersetup{draft}

\begin{document}
\label{firstpage}
\pagerange{\pageref{firstpage}--\pageref{lastpage}}
\maketitle

\begin{abstract}
Cosmological information from weak lensing surveys is maximised by sorting source galaxies into tomographic redshift subsamples. Any uncertainties on these redshift distributions must be correctly propagated into the cosmological results. We present \hyperrank, a new method for marginalising over redshift distribution uncertainties, using discrete samples from the space of all possible redshift distributions, improving over simple parameterized models. In \hyperrank the set of proposed redshift distributions is ranked according to a small (between one and four) number of summary values, which are then sampled along with other nuisance parameters and cosmological parameters in the Monte Carlo chain used for inference. This approach can be regarded as a general method for marginalising over discrete realisations of data vector variation with nuisance parameters, which can consequently be sampled separately from the main parameters of interest, allowing for increased computational efficiency. We focus on the case of weak lensing cosmic shear analyses and demonstrate our method using simulations made for the Dark Energy Survey (DES). We show the method can correctly and efficiently marginalise over a wide range of models for the redshift distribution uncertainty. Finally, we compare \hyperrank to the common mean-shifting method of marginalising over redshift uncertainty, validating that this simpler model is sufficient for use in the DES Year 3 cosmology results presented in companion papers.
\end{abstract}





\section{Introduction}
\label{section:introduction}
As photometric galaxy surveys begin to map large fractions of the sky at deeper magnitudes, stringent control of systematic errors and uncertainties is required to take full advantage of the statistical power of such surveys. Combining measurements of weak lensing and spatial clustering of distant galaxies (and cross-correlations of these two signals as galaxy-galaxy lensing) has steadily become a very competitive probe of the expansion history of the Universe and its constituents \citep[e.g.][]{2018PhRvD..98d3526A, 2019PASJ...71...43H, 2020PASJ...72...16H, 2020arXiv200715632H}. The Dark Energy Survey Year 3 \citep[DES-Y3][]{y3-3x2ptkp} results, of which this work forms a part, contains information from over 100 million galaxies.
One of the key required measurements in such analyses is the line of sight distribution of both the galaxies for which the shapes are measured (the \emph{source} sample) and the generally lower redshift galaxies used to trace the massive structures acting as lenses for the source sample (the \emph{lens} sample).
In an approach known as `tomography' \citep{1999ApJ...522L..21H} the source sample is sub-divided into different bins of distance, allowing us to further study the evolution of massive structure across cosmic time by observing how the lensing signal changes as a function of distance. Knowledge of the distance distribution to the source sample is a crucial ingredient in this. Of particular interest for modern cosmology, the statistical properties of dark matter structures as a function of cosmic time are a promising probe of dark energy.

Cosmological redshift $z$ is the observable most commonly used as a proxy for the distances to both 
galaxy samples but the methods to estimate distance via redshifts often suffer from limitations which make this one of the most difficult uncertainties to adequately model for the cosmological analysis. Estimating the redshift with high accuracy using spectroscopy is prohibitively expensive in telescope time for the large numbers of galaxies required for cosmology using weak lensing and suffers from selection effects caused by the incompleteness at fainter magnitudes \citep[e.g.][]{2020MNRAS.496.4769H}. Photometric redshift (\photoz) methods instead estimate the redshift based on measurements of fluxes in a number of photometric bands, and present a viable alternative in terms of sky and redshift coverage and completeness, but suffer from relatively much larger uncertainties given the highly degenerate problem of estimating $z$ based on wide band photometry.
A wide range of \photoz methods are used to estimate redshifts from band magnitudes; see \cite{2020MNRAS.499.1587S} and references therein for a recent review.

Current galaxy surveys rely on a combination of spectroscopic and  photometric redshifts, plus clustering patterns, to train, calibrate, and validate different methods. These methods can be broadly classified into three types, based on the information and ancillary data used to estimate redshift.
(i) Template fitting methods \citep[see Section 3.1 of][for a review]{2020MNRAS.499.1587S} which rely on finding the best-fit template redshift from 
a library of spectral energy distributions (SED) characterising a range of galaxy types.
(ii) Machine learning based techniques \citep[see Section 3.2 of][for a review]{2020MNRAS.499.1587S} which map the colour space into redshifts.
While the range of approaches used is fairly wide, the general idea consists of using a training set of secure redshifts obtained using either spectroscopy or large sets of narrow-band filter photometric observations to train the algorithm.
(iii) Using spatial correlation between galaxies and a set of tracers with secure redshift information to obtain additional constraints on redshift (often known as `clustering redshifts'). See the introductory sections of \cite*{y3-sourcewz}; \cite{y3-lenswz} for recent reviews.

Irrespective of the chosen method there will be an irreducible uncertainty in the galaxy distances arising from the finite number of photons received in each band, the widths of the bands and our limited knowledge of true galaxy spectral energy distributions. Where galaxies are observed only in a few ($\sim 1 - 10$) photometric bands there are also fundamental degeneracies where two galaxies at very different redshifts can produce identical observed data.
This uncertainty must be propagated through to cosmological constraints.
Galaxies are conventionally grouped into a small number ($\sim 5$ for current experiments) of tomographic redshift bins.
Cosmological observables of weak lensing, galaxy clustering and galaxy-galaxy lensing formed from each of these tomographic bin sub-samples are dependent on the number density distribution of the sources as a function of redshift within each bin, \nz.
If each individual galaxy's redshift were known with perfect precision and accuracy, these \nz would be non-overlapping, and their shapes would follow the true distribution in redshift of galaxies which are really in these bins.
However, in real cases where the one point summary statistic used for binning is noisy, biased, or both, the \nz within different tomographic bins acquire stretched tails which often overlap across the full redshift range of the survey.

In order to constrain cosmological parameters, expected weak lensing observables for a galaxy sample with the estimated \nz and in a given cosmology are computed and compared with the data.
Monte Carlo methods are then used to map the posterior for cosmological model parameters and hence constrain our physical model for the Universe.
In this inference process, uncertainties on the measured \nz for each tomographic bin are marginalised over, typically widening the uncertainties on the cosmological parameters of interest.
Incorrectly quantifying the uncertainty on the \nz or incorrectly marginalising over it can significantly affect cosmological parameter estimation and model selection.
Indeed \cite{2020A&A...638L...1J} have argued that the adoption of different models for calibration of redshift distributions and their uncertainties for weak lensing experiments can explain the observed apparently significant difference in cosmological parameters between different weak lensing experiments and Cosmic Microwave Background (CMB) experiments.

In this paper we introduce \hyperrank, a new method which allows uncertainties in galaxy redshift distributions \nz to be propagated into Monte Carlo chains generating cosmological results. \hyperrank takes as input a finite set of samples of \nz\ drawn from the distribution implied by the redshift calibration process.  It maps these onto a low-dimensional space of continuous variables which the cosmology sampler can treat as free parameters.  We test that \hyperrank does this both \emph{correctly}, in that the allowed uncertainty is fully explored, and \emph{efficiently}, in that fewer likelihood evaluations are computed than in the case where an arbitrary choice of \nz realisation is made at each step. This approach allows for the inclusion of a much wider range of types of uncertainty on \nz to be used in cosmological inference than have been included in the majority of previous analyses.

In \cref{section:y3-discrete-redshifts} we review methods of quantifying uncertainties on the redshift distributions of galaxy samples used for cosmology, motivating the introduction of the new \hyperrank method, which is then described in \cref{section:hyperrank}, in both the simplest one-dimensional case and an extended multi-dimensional case. In \cref{section:simulation} we then perform tests of the performance of \hyperrank on a simulated version of the DES-Y3 experiment. In \cref{sec:correct-exploration} we verify that in cases where redshift distribution uncertainty is known, \hyperrank correctly marginalises over this uncertainty, for four representative models of the uncertainty. In \cref{subsection:simulation-efficiency} we also show that the use of \hyperrank to explore the uncertainties results in better performance (in terms of fewer numbers of Monte Carlo steps required) than random, un-ranked exploration of realisations of possible redshift distributions. We also explore the performance of a number of different choices of variables on which to perform the ranking and find, for our fiducial case, the number of discrete samples from the possible redshift distributions which are required for the cosmological results to converge to those of a known case where continuous sampling is possible. \Cref{section:y3} describes the application of \hyperrank to the real DES-Y3 data, with the results presented in \cite{y3-cosmicshear1}. Finally in \cref{section:conclusions} we discuss our results and conclude.


\section{Marginalisation of redshift uncertainty}
\label{section:y3-discrete-redshifts}

\begin{figure}
    \centering
    \includegraphics[width=\columnwidth]{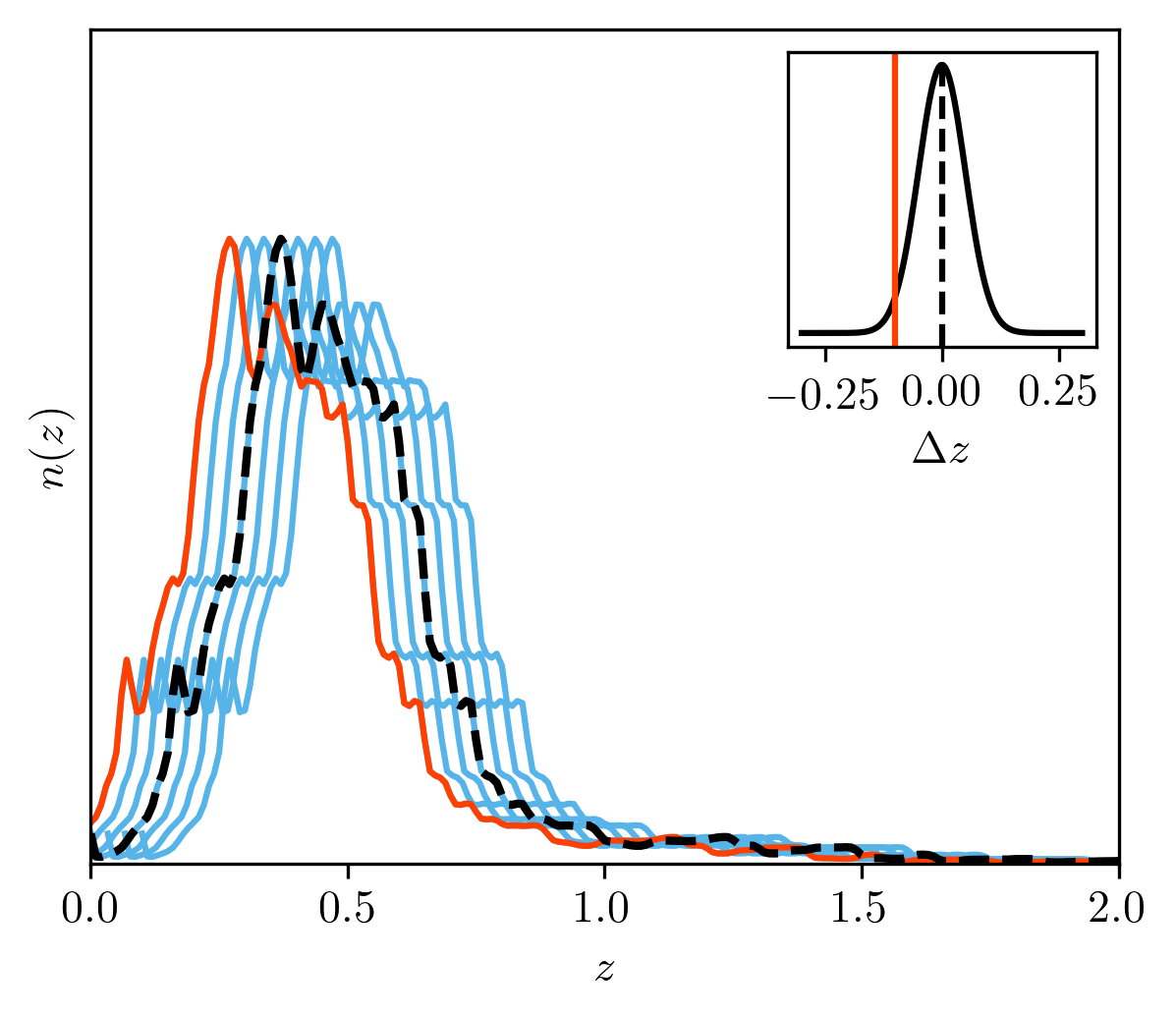}\\
    \includegraphics[width=\columnwidth]{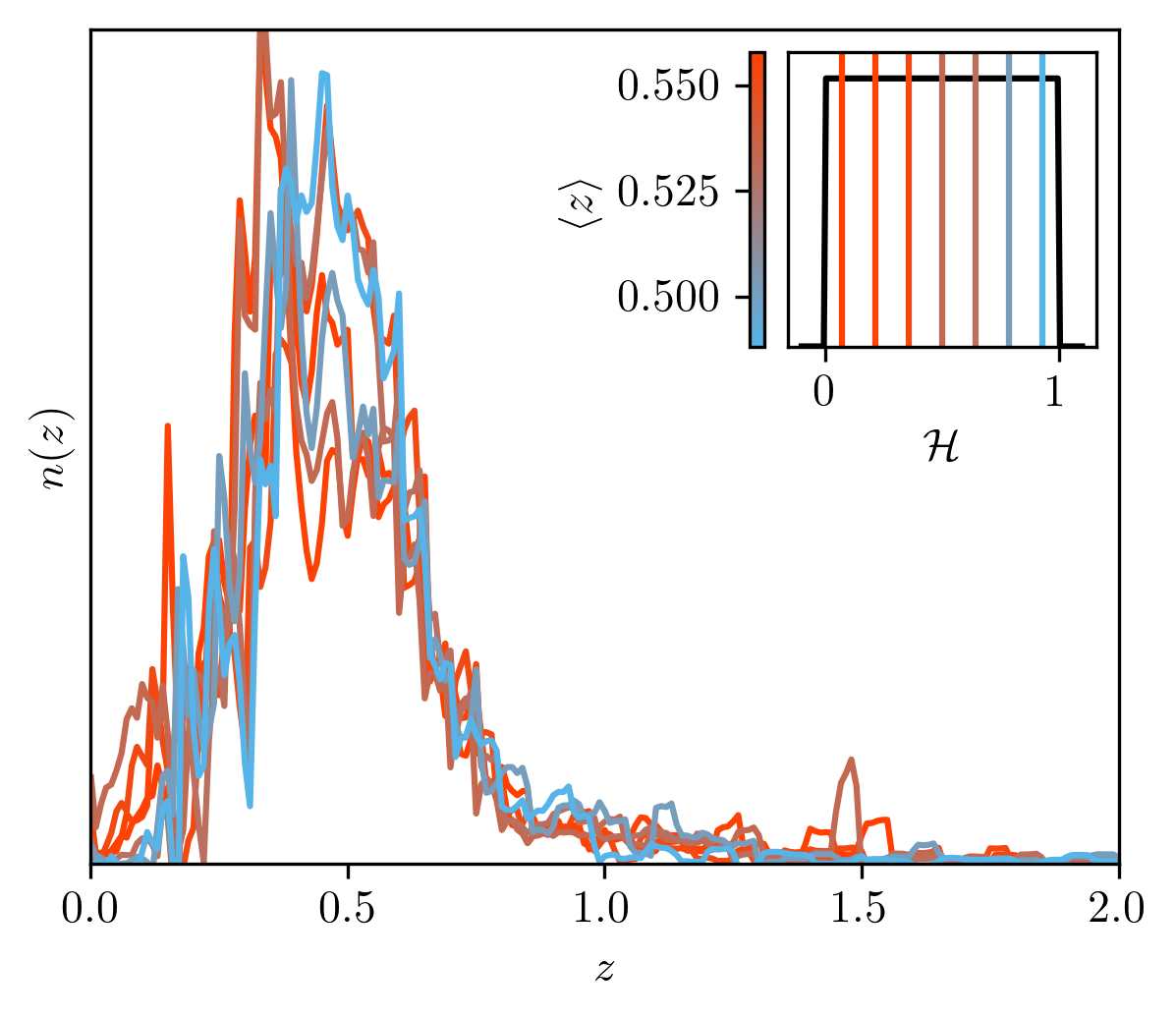}
    \caption{\emph{Upper}: The $\Delta z$ marginalisation scheme, where a fiducial redshift distribution (black dashed) is shifted horizontally at each Monte Carlo step by a value drawn from a Gaussian distribution (inset, with draw from the $2\sigma$ tail highlighted in red).
    \emph{Lower}: Discrete realisations of possible \nz are shown with colours corresponding to the mean redshift of each realisation $\langle z \rangle$, which can be mapped to a ranking hyper-parameter $\mathcal{H}$ which is then marginalised over on the Monte Carlo chain. Inset shows the uniform distribution for $\mathcal{H}$ which is sampled from, and the centres of the regions corresponding to each coloured \nz realisation.
    }
    \label{fig:nz-deltaz}
\end{figure}

In general, for an inference problem in which we have a model containing parameters of interest $\theta$ (such as the cosmological parameters) and a set of nuisance parameters $\alpha$ (such as parameters relating to redshift distribution uncertainty), we form posterior probability distributions:
\begin{equation}
\label{eq:bayes}
    P(\mathbf{\theta}, \alpha | \mathbf{x}) \propto \mathcal{L}(\mathbf{x}|\theta,\alpha)P(\theta, \alpha)
\end{equation}
where $\mathcal{L}(\mathbf{x}|\theta, \alpha)$ is the likelihood function for the data $\mathbf{x}$ and $P(\theta, \alpha)$ is a prior probability distribution.
When generating samples from the posterior with a Monte Carlo process, the nuisance parameters are typically sampled jointly with the parameters of interest and then marginalised over, providing a marginal posterior on the model parameters $\theta$ in which the uncertainty on $\alpha$ is accounted for.
In the particular case of redshift distributions in cosmology analyses, a common approach is to provide a fiducial tomographic redshift distribution and characterise its uncertainty using the nuisance parameter of a shift \deltaz along the $z-$axis. A different parameter $\deltaz_i$ is used for each tomographic bin, with each drawn from a Gaussian prior informed by observations and/or simulations.
This approach is depicted in the upper panel of \cref{fig:nz-deltaz} and has been used in DES SV \citep{2016PhRvD..94d2005B},  DES Y1 \citep{2018MNRAS.478..592H}, HSC \citep{2019PASJ...71...43H} and KiDS-1000 \citep{2020arXiv200701844J}.
However, whilst convenient and capturing the uncertainty in the mean of the redshift distributions, which is strongly correlated with cosmology, it is not physically well motivated and severely restricts the possible functional forms which a proposed \nz may take.

In contrast to the \deltaz approach, we may wish to consider alternatives which allow for a much wider range of uncertainty in the functional forms of the \nz.
It is possible to take a simulations-based approach, in which realisations for the possible \nz of a survey are generated by multiple realisations of mock versions of the survey created from independent patches of cosmological simulations. 
Alternatively, we may explicitly parameterise the \nz as a set of histogram bin heights $n(z_i)$ which give the counts of sources within a small redshift interval and try to infer these quantities from the data.
This approach creates principled models of the joint probability distribution function for all of these bin heights given the photometric data available on the observed galaxies.
This is most readily done as a Bayesian Hierarchical Model and has been recently advocated in \cite{2016MNRAS.460.4258L}, \citet{2019MNRAS.483.2801S} and \cite{2020MNRAS.491.4768R}.
Outputs from this procedure are samples from the joint posterior for all of the histogram bin heights which together make up the full shape of the \nz.
Each sample consists of a possible realisation of what the full \nz could look like, discretised as $n(z_i)$.
An ideal approach would be to treat each of these $n(z_i)$ as a model parameter and jointly infer them with the cosmological model parameters before marginalisation.
In reality this is impractical; the redshift resolution required to capture important features of the model which impact cosmological inference but are not convolved with broad redshift kernels, such as intrinsic alignments, would demand  hundreds of additional nuisance parameters.
Current implementations of galaxy survey analysis pipelines \citep[such as that in \cosmosis used for DES][]{2015A&C....12...45Z} typically take $\sim1-10$ seconds per likelihood evaluation, meaning the addition of hundreds of parameters would mean the samplers used (MCMC such as \emcee \citealt{2013PASP..125..306F} or nested sampling such as \multinest \citealt{2009MNRAS.398.1601F} or \polychord \citealt{2015MNRAS.450L..61H}) could not map the full posterior in a timely manner. It should be noted that \cite{2017MNRAS.465.1454H} were able to run 750 MCMC chains in order to use a different bootstrap resampling realisation of their \nz each time, before combining these chains; we do not expect this to be feasible for the DES-Y3 pipeline. 
\response{Other methods have also been proposed to address the uncertainty associated to large number of nuisance parameters, including Gaussian mixture models, which are flexible and may be analytically marginalised over \citep{2020JCAP...10..056H, 2020arXiv201207707S}, and the use of flat or Gaussian priors to characterise variations to sets of arbitrary functions used to evaluate a Gaussian likelihood \citep{2010MNRAS.408..865T, 2011MNRAS.410.1677K}}

Here we consider an alternative approach in which the set of samples from the \nz posterior, \response{each consisting of a collection of histogram values for each tomographic bin,} are generated outside (before) the cosmological parameter inference Markov chain.
This set of realisations can then be used by choosing a new \nz in every likelihood evaluation within the cosmological parameter inference chain, allowing higher-order modes of uncertainty in \nz to be propagated into cosmological parameter constraints.
\response{It is important to note that the \nz realisations are drawn simultaneously for all tomographic bins, which also allows the propagation of uncertainty originating from correlations between tomographic bins.}

A way of performing this analysis would be to randomly sample a different redshift distribution on each likelihood evaluation within the Monte Carlo chain.
This has potential negative effects on the behaviour of Monte Carlo samplers which rely on the posterior function being a smooth function of the sampled parameters.
A random approach can break the smoothness of the likelihood (as shown in \cref{fig:1d-logposterior}) in the other parameter dimensions leading to unnecessarily high sample rejection rates, requiring large number of likelihood evaluations for convergence and potentially disrupting convergence criteria for different samplers.

Here, we present \hyperrank, a way to overcome these computational limitations, whilst still exploring the space of uncertainty available from the discrete \nz realisations. In \hyperrank we construct a mapping between the index of an ordered set of \nz realisations and a continuous parameter $\mathcal{H}$, such that the likelihood function $\mathcal{L}(\theta, \mathcal{H})$ is smooth on this new space and the prior $P(\mathcal{H})$ preserves an equal weighting of the \nz samples through assigning them to an evenly-spaced grid.


\section{The hyper-ranking method}
\label{section:hyperrank}

After a discrete set of realisations of tomographic bin redshift distributions $n_i(z)$ have been generated, we wish to correctly and efficiently marginalise over the uncertainty embodied by them, within a cosmological parameter inference Monte Carlo chain.
We introduce the idea of \hyperrank-ing in which the full set of realisations is mapped onto a small (in this work between one and four) number of parameters $\mathcal{H}$.
The \nz realisations are ordered according to a set of descriptive values $\mathbf{d}$ which are \emph{a priori} expected to correlate strongly with values of the cosmological parameters of interest. 
\response{This ordering preserves the tomographic nature of each realisation, meaning the sampling stage selects the set of all tomographic bins' distributions simultaneously, without mixing different realisations.}
The rank parameters $\mathcal{H}$ become the nuisance parameters which are sampled (and subsequently marginalised over) in the cosmological analysis.
Choosing descriptive values $\mathbf{d}$ which correlate with the cosmological parameters of interest ensures the likelihood varies as smoothly as possible along each dimension of the rank parameters. The ranking parameters $\mathcal{H}_j=\mathcal{H}(\alpha_j)$ must also be such that realisations with similar descriptive values are mapped close to each other. Furthermore, the $\mathcal{H}_j$ must be such that a uniform prior on $\mathcal{H}$ preserves equal probability on all input \nz samples. We consider the cases below first in which we have one ranking parameter and then multiple ranking parameters.
\response{We choose to mainly use the mean redshift $\langle z \rangle$ and mean inverse comoving distance $\langle 1/\chi \rangle$ of each tomographic bin as descriptive values $\mathbf{d}$ here, but emphasise the \hyperrank method is not limited to these two options only. We expect the choice of ranking method to only affect sampling efficiency and not the inferred parameter contours.}

\subsection{One dimensional case}
\label{subsection:hyperrank-oned}
We initially consider the case in which a single \hyperrank parameter is used to rank all realisations.
Since the mean redshift of the distribution \nz varies the overall amplitude of lensing expected for a given source galaxy sample, it is expected to correlate with the cosmological parameters of interest (here, the matter amplitude parameter $S_8$). We therefore consider a basic \hyperrank approach in which there is only one descriptive parameter per realisation of the full \nz  and it is based on the weighted mean redshift of a combination of tomographic bins,
\begin{equation}
    \mathbf{d} = \frac{\sum w_i \langle z \rangle_i}{\sum w_i},
    \label{eq:binmeanz}
\end{equation}
where $i$ is the index of each tomographic bin and $w_i$ is the corresponding weight, which can embody (for instance) the number of assigned galaxies to each tomographic bin.
The \nz realisations are then ranked according to their descriptive value $\mathbf{d}$ and mapped to a continuous hyper-parameter $\mathcal{H} \in [0,1)$, which is then sampled in the Monte Carlo chain. Each sampled value of $\mathcal{H}$ corresponds to a stored \nz realisation which is then used in the likelihood evaluation.
This approach is demonstrated in the lower panel of \Cref{fig:nz-deltaz} which shows a small sample of \nz realisations coloured according to their mean redshift and assigned a range of $\mathcal{H}$ values depending on their ranked position.

An alternative set of descriptive values are the mean inverse comoving distance of sources, $\langle 1/\chi \rangle$. The correlation of this quantity with cosmological posterior value can be motivated by its relation to the lensing efficiency functions used in the calculation of the shear power spectrum, which can be written as,
\begin{equation}\label{eq:P_kappa}
	P_{\kappa}(\ell)
	= \frac{9 H_0^4 \Omega_m^2}{4 c^4} \int_{0}^{\chi_\Hor} \! g^2(\chi) \, \frac{P_{\delta}(\ell/\chi; \chi)}{a^2(\chi)} \, \D\chi \;,
\end{equation}
where $\chi_\Hor$, $a(\chi)$ and $P_\delta$ are the comoving horizon, scale factor and matter power spectrum, respectively and the lensing efficiency $g(\chi)$ at comoving distance $\chi$ is defined as:
\begin{equation}\label{eq:g}
	g(\chi)
	= \int_{\chi}^{\chi_\Hor} \! n(\chi') \, \frac{\chi' - \chi}{\chi'} \, \D\chi' \;,
\end{equation}
and depends on the comoving distance distribution $n(\chi)$ of sources, or equivalently their redshift distribution \nz.
By evaluating at $\chi = 0$ and differentiating the above definition for the lensing efficiency we obtain
\begin{eqnarray}\label{eq:gprime}
    \left. g(\chi)\right|_{\chi = 0} = 1\\
    \left. g'(\chi) \right|_{\chi = 0} = -\langle 1/\chi \rangle_{n},
\end{eqnarray}
where $g'(\chi) = \mathrm{d} g / \mathrm{d} \chi$, which are boundary conditions for the lensing efficiency functions hence control their overall shape. See \cite{2020OJAp....3E...6T} for discussions of the importance of constraining $g'(\chi)$ in weak lensing studies.

\begin{figure}
    \centering
    \includegraphics[width=\columnwidth]{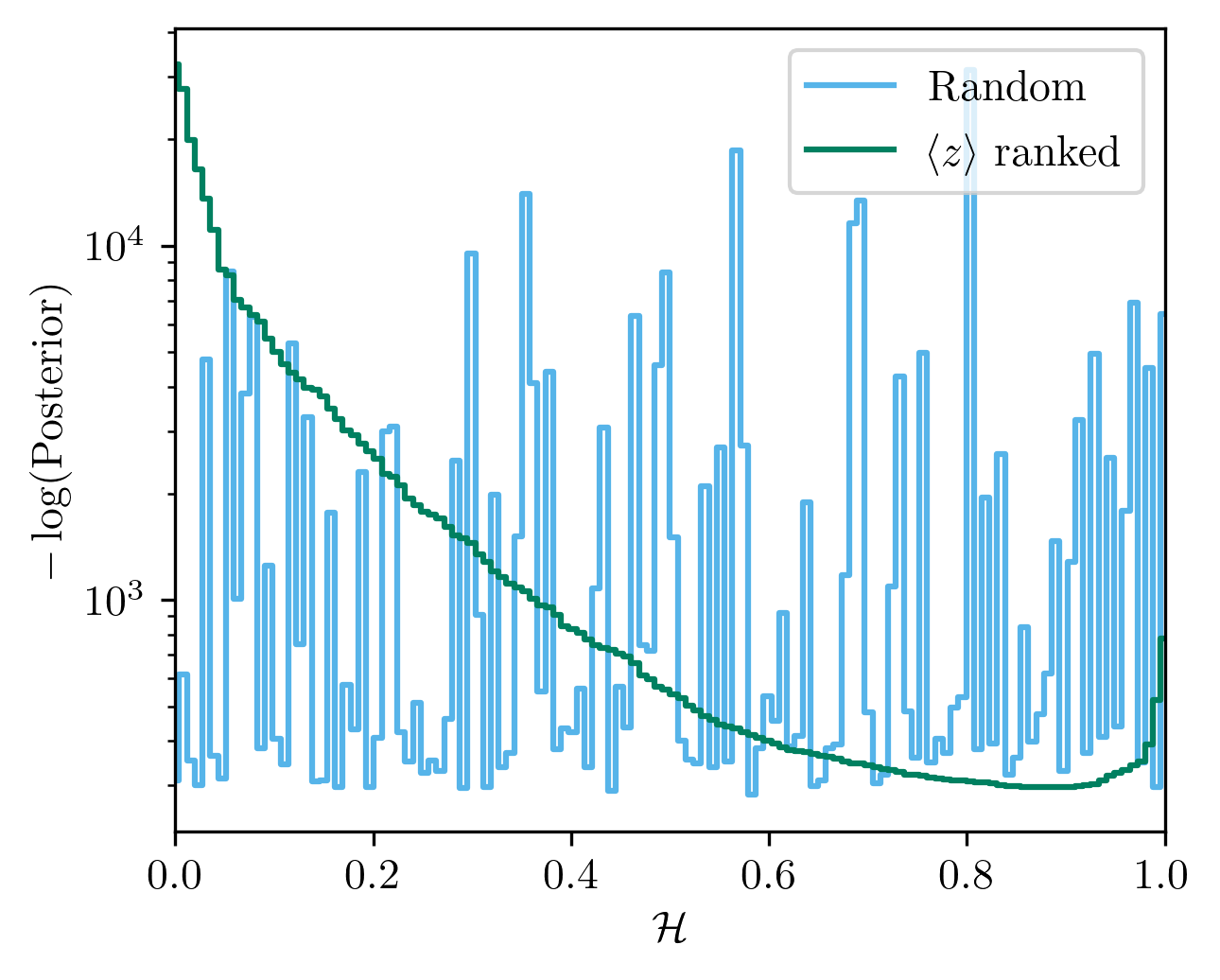}
    \caption{The log posterior for the \hyperrank parameter of a single tomographic bin when holding all other parameters fixed, contrasting the cases of random ranking (which gives no smooth posterior for the sampler to explore) with mean redshift ranking (which does give a smooth posterior surface).}
    \label{fig:1d-logposterior}
\end{figure}

The mapping of distributions is not invariant to the choice of ordering being the mean redshift or the mean inverse comoving distance. In one dimension, both are examples of ranking parameters capable of providing the smooth likelihood necessary for efficient mapping of the posterior (as can be seen in Figure \ref{fig:1d-logposterior}), as well as correctly including the space of uncertainty spanned by the provided set of \nz realisations. 
In \cref{section:simulation} below we consider only the mean redshift ranking for the one dimensional case, but observed comparable performance for the inverse comoving distance ranking in our tests.

\subsection{Multi-dimensional case}
\label{subsection:hyperrank-multid}

While the one-dimensional approach presents a clean and simple strategy to arrange and select realisations for each likelihood evaluation, it doesn't prevent cases where two realisations with very different descriptive values are assigned a similar rank---e.g. two realisations have very distinct $\langle z \rangle_i$ in individual redshift bins, but similar when averaged over bins as per \cref{eq:binmeanz}. \response{Indeed in our initial tests with DES-Y3 simulations it was found this was often the case, leading to realisations ranked closely by a single mean redshift parameter having significantly different posterior values, hence leading to poor efficiency in the cosmology chains.}
\response{To address this} we describe a generalisation to rank distributions using multiple dimensions, which allows to use more than one descriptive parameter $\mathbf{d}$ to assign the proposal \nz realisations to a space of hyper-parameters $\mathcal{H}$.
\response{Matching of the number of descriptive values and the dimensionality of the redshift distributions (e.g. number of tomographic bins) is not a requirement, and we find here the best performance is achieved when this is not the case.}

Each of the $N_p$ proposals for \nz is assigned a position in a uniform multi-dimensional grid, $\mathbf{u}$, according to a set of $N_d$ descriptive values $\mathbf{d} = {d_1, ... , d_{N_d}}$.
This grid is contained inside a $N_d-$dimensional unit hyper-cube, and the continuous parameters $\mathcal{H}_j \in [0,1)^{N_d}$  are sampled in the Monte Carlo chain.
For each $\mathcal{H}$ value chosen by the sampler, the method returns the closest $\mathcal{H}_i$ in the grid, which has been assigned to one of the $N_p$ \nz realisations.

\begin{figure*}
    \centering
    \includegraphics[width=\textwidth]{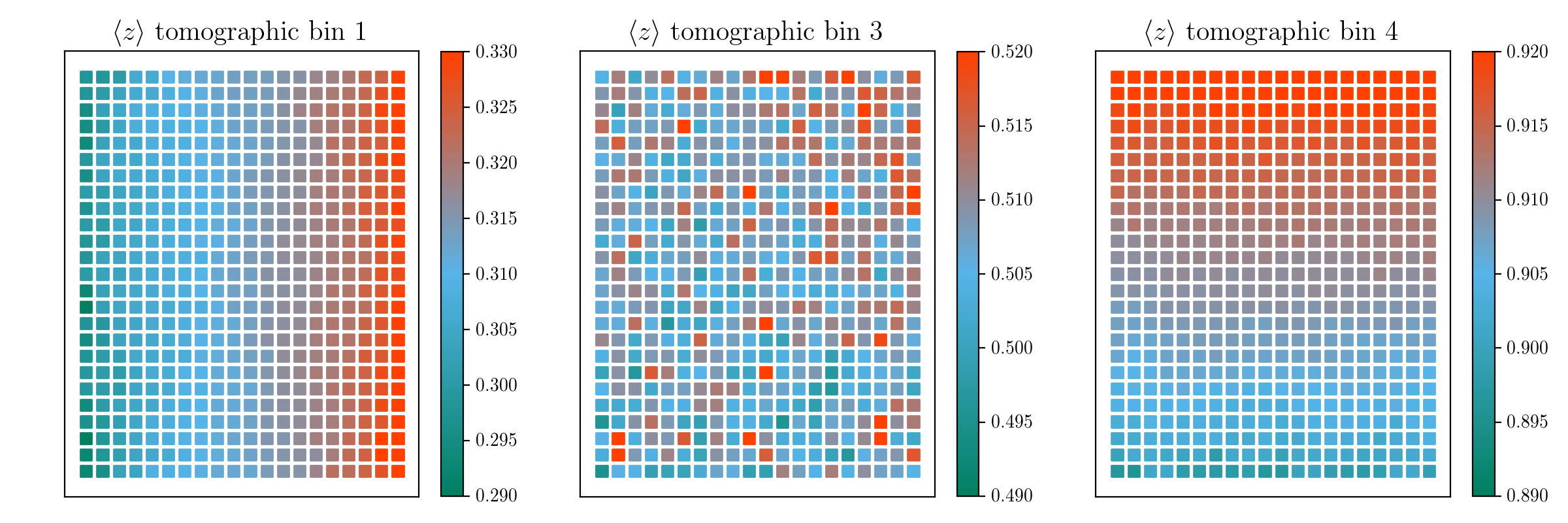}
    \caption{$25\times 20$ ranking map generated using the mean redshift for tomographic bins 1, 2 and 4 from a set of 500 redshift distributions.
    Panels show distributions located in the same positions, but the color scale shows the mean redshift for the corresponding tomographic bin.
    It can be seen that the mapping scheme permits realisations to remain close to other realisations with similar descriptive values used for the mapping, and has a smooth variation in the directions of the hyperparameters mapped to each dimension of the grid.
    The arrangement does not necessarily result in a smooth ordering of all tomographic bins, as can be seen from the middle panel where the mean redshift from a bin not used of the mapping is displayed.}
    \label{fig:2D_map_assignment_buzzard}
\end{figure*}

We now need to consider how to preserve the notion of ordering the set of \nz by descriptive values in this multi-dimensional space, preserving the notion of a `neighbourhood' where realisations with similar descriptive properties are grouped close together. One approach to find the optimal relative positions is to use the solution to the \emph{Linear Sum Assignment Problem} \citep[e.g.][]{Burkard1980}.
Given a set of $N_p$ workers (points in the descriptive value space) we want to find an assignment to $N_p$ fixed jobs (i.e. fixed grid positions in the unit hyper-cube) such that the sum of the cost to assign each worker to one and only one job (the distance from descriptive value space to hyper-cube position) is minimised:
$$\min \sum C_{ij} X_{ij},$$
where $C_{ij}$ is the cost matrix of assigning each sample $\mathbf{d_i}$ to each point $\mathbf{u}_j$ of the grid, and $X_{ij}$ is a binary matrix indicating which position is assigned to each set of descriptive values.
If we use an Euclidean distance metric such that $C_{ij} = |\mathbf{d_i} - \mathbf{u_j}|^2$, the resultant assignment minimises the total distance moved by the points to the positions on the grid ensuring that any notion of neighbourhood between points in the original space of descriptive parameters is preserved in their new unit hyper-cube grid positions.  We implement this technique by first linearly rescaling the $\mathbf{d}_i$ so that they span a unit hypercube.
Figure \ref{fig:2D_map_assignment_buzzard} shows the resultant 2-dimensional assignment for a set of 500 realisations, each comprised of a set of four tomographic bins, using as descriptive parameters the mean redshifts of tomographic bins 1 and 4, arranged in a $25 \times 20$ map.
\response{Because of the finite number of available realisations the use of additional dimensions can quickly have the undesired effect of reducing the amount of realisations available with which to fill each direction of the multidimensional grid.
This can result in the exacerbation of the convergence problem, with few available samples creating large jumps in posterior as a function of the $\mathcal{H}$ parameters. For example with 4096 realisations, double the grid size is available with $N_d = 3$ dimensions compared to $N_d = 4$.}

In the case of $N_d = 1$, where a single characteristic value describes each realisation and the arrangement of points is done over a grid in the interval $[0,1)$, the optimal distribution is the one which ranks the points in order, corresponding to the case described in \Cref{subsection:hyperrank-oned}.
Analogous to the one dimensional case, we propose the use of mean redshift $\langle z \rangle$ or mean inverse comoving distance $\langle 1/\chi \rangle$ of the individual tomographic bins as sources of descriptive values to map the realisations to the hyper-cube.

Ideally, the dimensionality $N_d$ of the hyper-ranked space is low enough to maintain an efficient cosmological sampler, but high enough that the variation in the log posterior probability from Eq.~(\ref{eq:bayes}) in small regions of $\mathcal{H}$ is $\ll1.$ This would allow any sampling process to smoothly traverse the full space of all \nz variations that influence the parameters of interest.

We can optimize the reduction of the nuisance-variable vector $\alpha$ (e.g. all of the freedom of \nz) into a lower dimensional hyperspace by using the Karhunen-Lo\`eve (KL) transformation.  When the observational data vector $D$ has a Gaussian likelihood with covariance matrix $C_D$ and mean value $\hat D(\theta,\alpha)$, we find the eigenvectors $e^k$ of the matrix
\begin{equation}
    \left(\frac{\partial D}{\partial\alpha}\right)^T C_D^{-1} \left(\frac{\partial D}{\partial\alpha}\right).
    \label{eq:klt}
\end{equation}
where the derivatives are taken about some reference values of $\theta$ and $\alpha.$  The best choice of \hyperrank descriptive values $(d_1,d_2,\ldots,d_K)$ will be to order the eigenvectors by decreasing eigenvalues, and assign $d_k=\alpha e^k$ for each input sample. Successive $d_k$ values have decreasing influence on the cosmological model.
The sum of the eigenvalues at $k>K$ then describes the ``roughness" of the log-posterior in the $\mathcal{H}$ space.
Using this principal component analysis (PCA)-style approach, we can choose the first $K$ components of the decomposition as descriptive values to inform the ranking map and assign each component to one \hyperrank parameter each.

The main caveat is that this approach defines a set of descriptive values which are optimal only near the reference cosmology chosen to compute the KL components.
While ideally one would want to use a large number of dimensions to help construct a space where the posterior is as smooth as possible, this comes at the expense of having to construct a grid with a low number of points per dimension, if the number of input samples of \nz is held fixed.
This can result in a noisy posterior as a function of the hyper-parameter $\mathcal{H}$ if a given dimension of $\mathcal{H}$ is sparsely sampled and has large steps between samples.
While a large number of realisations can help construct a grid with a reasonably large number of realisations per side of the grid, the method to solve the linear sum assignment problem scales as $\mathcal{O}(N_p^3)$, which quickly becomes unmanageable.
In \cref{subsection:simulation-efficiency} we explore the effects the dimension of the ranking and choice of descriptive value have on sampling efficiency, testing the mean redshift, inverse comoving distance, and KL approaches with 3 components each.

\section{Tests on simulations}
\label{section:simulation}

We now test the \hyperrank method for marginalising over redshift distribution uncertainty and explore its configuration, with the target of using it for the weak lensing source redshift distributions in the DES Year 3 cosmological analysis. We investigate the \hyperrank method's ability to marginalise over the \nz uncertainty:
\begin{itemize}
    \item \emph{correctly}, in that it proportionately explores the space of possible \nz represented by the discrete realisations which are provided as an input.
    \item \emph{efficiently}, in that as few likelihood evaluations as practically possible are required before the Monte Carlo process converges to the posterior.
\end{itemize}
We test the correctness by comparing the recovered posteriors on the $S_8 = \sigma_8 \sqrt{\Omega_m / 0.3}$ cosmological parameter obtained from a cosmological inference pipeline.
We generate sets of \nz samples using a number of well defined procedures in which the method for generating realisations involves drawing a \deltaz shift from a known analytic distribution. We then run analyses using \hyperrank to marginalise over these uncertainties and compare the results to a set of chains in which the known analytic distributions from which the \deltaz were drawn are used again to marginalise over the nuisance parameter. Hence we verify that, in the case where discrete samples represent a model for uncertainty on \nz, the use of \hyperrank correctly explores this uncertainty. The tests show that \hyperrank is capable of correctly marginalising over redshift distribution uncertainties in cases where a correct and simple model for them is known, without making assumptions on the form of the uncertainty model. This model-agnosticism represents an advantage in the case of real experiments, where it may not be known \emph{a priori} if one or any of the simple models is adequate for obtaining small, unbiased posteriors.

We also compare the results from analyses using \hyperrank to ones in which discrete \nz realisations but no ranking (or equivalently random ranking) are used, showing that the imposition of the \hyperrank ranking does not bias or unduly constrain the cosmological parameter space explored.

We test the improvement in computational efficiency gained from using \hyperrank by comparing one- and multi-dimensional implementations of \hyperrank to a mode in which no ranking is performed and at each likelihood evaluation an \nz is chosen from the available realisations at random.

Finally, we also test the convergence of \hyperrank for the configuration required for DES-Y3 cosmology, finding the number of \nz realisations which are required before systematic errors on the cosmology parameters from the discreteness introduced by \hyperrank become negligible.

Throughout these tests we use the DES-Y3 modelling choices, likelihood and pipeline software and configuration, which are described in detail in \citet{y3-cosmicshear1}; \citet*{y3-cosmicshear2}.
We only consider cosmic shear in our data vector, which reduces the dimensionality of the space of parameters to be sampled in the MC inference and enhances the effect of redshift systematics in the source sample.
Nevertheless, this method can be applied when using cosmic shear in a full 3x2pt analysis including galaxy clustering and galaxy-galaxy lensing and can also be used to marginalise over systematic uncertainties of the \emph{lens} in addition to the \emph{source} samples described here.

\subsection{Generation of fiducial redshift distribution}
Here we briefly describe the method by which the cosmic shear data vector and fiducial \nz used in our tests were generated. The methodologies and simulations are described in detail in \citet*{y3-sompz}, \citet*{y3-sourcewz} and \citet{y3-simvalidation}.
\subsubsection{Buzzard simulation}
\label{subsection:simulation-buzzard}

\begin{figure*}
    \centering
    \includegraphics[height=6cm]{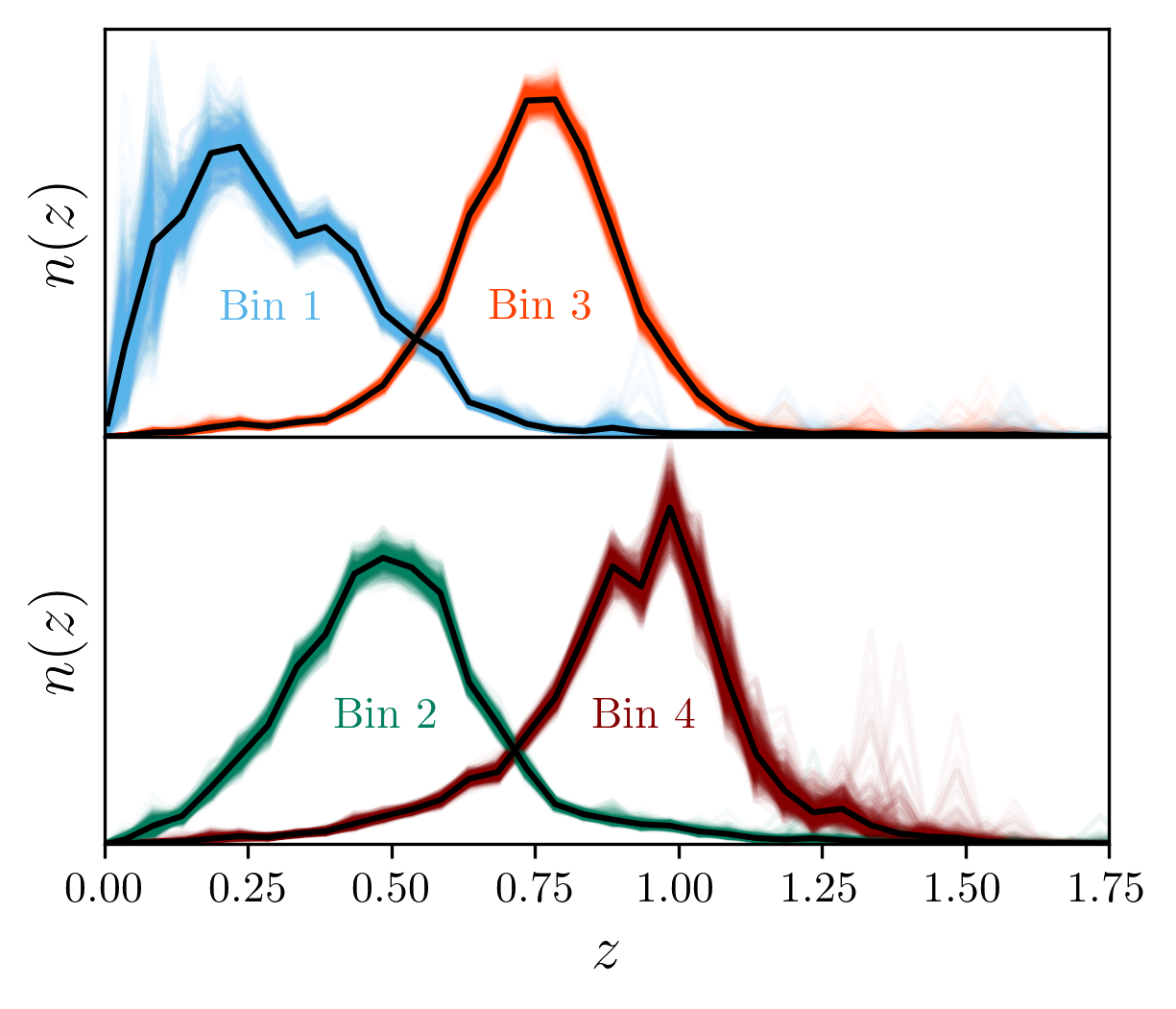}
    \hspace{-0.5cm}
    \includegraphics[height=6cm]{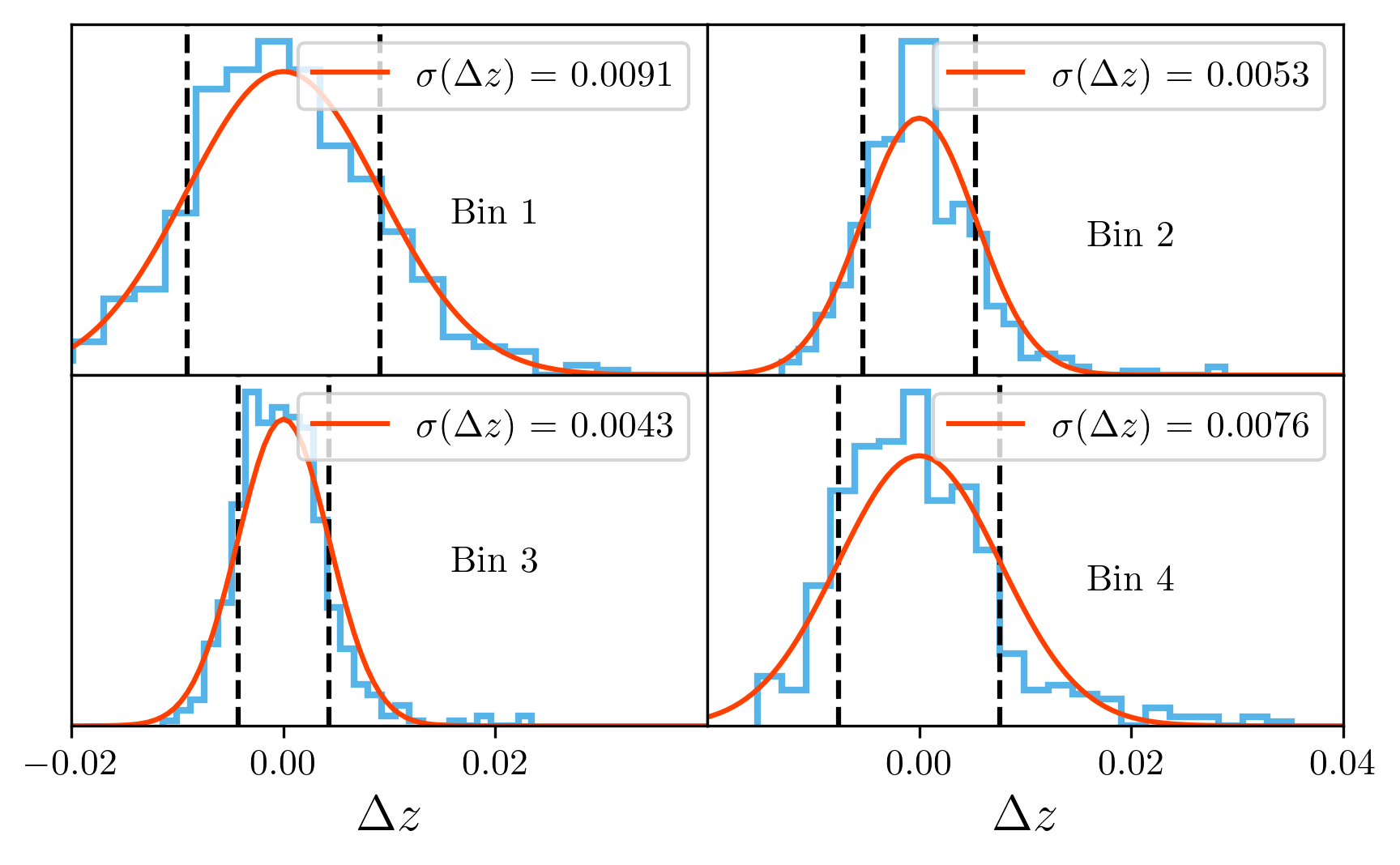}
    \caption{
    The \buzzard redshift distributions. \emph{Left:} The black lines show the redshift distribution \nzfid for each of the four redshift bins, averaged over all realisations.
    The light blue, red, green and brown lines show the full set of realisations for redshift bins 1 through 4, respectively, depicting the potential differences between independent samples of \nz posterior, and their peculiarities at the histogram level.
    \emph{Right:} Histogram of mean redshift for each of the four tomographic bins, computed from the ensemble of distributions on the left panel.
    Solid orange line traces the Gaussian fit to the histogram, described by the width \sigmadeltaz above each panel.
    }
    \label{fig:cosmos_buzzard_realisations}
\end{figure*}

The \buzzard simulations \citep[][]{y3-simvalidation} are a set of mock DES-Y3 surveys created from a suite of dark-matter N-body simulations using a memory-optimised version of L-GADGET2 \citep{2005MNRAS.364.1105S}.
Galaxies and their main morphological properties are added using \textsc{ADDGALS} \citep{2019arXiv190102401D}, matching projected clustering statistics and color-magnitude relations observed in the Sloan Digital Sky Survey Main Galaxy Samples (SDSS MGS as described in \citealt{2005AJ....129.2562B, 2009ApJS..182..543A}).
DES \textit{ugriz} and VISTA \textit{JHK} photometry is obtained from the simulated spectral energy distributions generated by \textsc{ADDGALS}.

\subsubsection{\sompz redshift distributions}
\label{sec:sompz}
The simulated photometry catalogues from \buzzard constitute the primary dataset to construct the fiducial \nz for our tests, using the \sompz method \citep*[fully described in ][]{y3-sompz}. This method makes use of three sets of observations: the full DES-Y3 wide field sample, the DES-Y3 Deep Fields \citep*{y3-deepfields} sample, and compilation of spectroscopic redshift surveys. Galaxies from the wide sample are grouped into \emph{phenotypes} using the Self-Organised Maps (SOM) method of dimensional reduction (see e.g. \citealt*[][]{y3-sompz}; \citealt{2020A&A...640L..14W,2015ApJ...813...53M}). The \balrog\ machinery \citep[which injects synthetic sources into DES data and recovers their properties, see][]{y3-balrog} is then used to quantify the probability of a given Deep Fields galaxy appearing to have a given phenotype when observed in the wide field. A second SOM dimensional reduction is then applied to the Deep Fields galaxy observations, with the spectroscopic sample used to characterise the true redshift distribution for each deep phenotype. In this way, information can effectively pass from the small, limited spectroscopic sample to the much larger wide sample through the intermediary of the deep sample.

In addition to this method of creating a best-estimate fiducial redshift distribution, we further consider realisations of possible \nz inferred from the simulated data using the method of \cite*{y3-sompz}; \cite{2020MNRAS.498.2984S,2019MNRAS.483.2801S}. This applies a three-step Dirichlet (3sDir) sampling to model the uncertainties on \nz histogram bin heights from sources including shot noise, sample variance, photometric calibration uncertainty, and method errors.
We use a set of 500 realisations generated this way, noting that samples are drawn jointly for all four tomographic redshift bins. The resulting estimated redshift distributions for \buzzard are shown as the coloured lines in the left panel of \cref{fig:cosmos_buzzard_realisations}. The fiducial realisation \nzfid is obtained from averaging the 500 realisations at the histogram level and re-normalising, and are shown as the black solid lines in the left panel of \cref{fig:cosmos_buzzard_realisations}.

\subsection{Exploration of uncertainties}
\label{sec:correct-exploration}

As a supplement to these full \sompz $+$ 3sDir realisations of the \buzzard \nz for testing, we also now take the fiducial \nz and construct sets of realisations of potential \nz using simple parametric models for the uncertainty.
We use analytic distributions to generate sets of mean redshift shifts \deltaz for each uncertainty model. We then compare the posteriors on cosmological parameters (and the effective \deltaz nuisance parameters) recovered by two chains:
\begin{itemize}
    \item a chain in which \hyperrank takes these realisations as an input set of proposed \nz
    \item a chain with \deltaz nuisance parameter marginalisation, using as a likelihood the same analytic distribution which was used to generate the realisations
\end{itemize}
To perform our sampling we use the \multinest sampler, with 500 live points, \texttt{tolerance} = 0.3, and \texttt{efficiency} = 0.01. We follow the setup for the DES-Y3 cosmic shear analysis described by \cite{y3-cosmicshear1}; \citet*{y3-cosmicshear2} in terms of angular scale cuts, tomographic redshift binning and modelling choices and marginalisation over other nuisance parameters such as shear calibration biases or Intrinsic Alignment model parameters. In most of the tests, and unless explicitly noted, we use the default three dimensional \hyperrank configuration described at the start of section \ref{section:y3}.

We will describe each test in the following section, as well as the results for each one presented in \cref{fig:results-gaussian,fig:results_nongaussian,fig:results_corrgaussian,fig:results_amplified}. In each of these figures, the top panels show the one dimensional posterior constraints recovered on $S_8$ and the means of the redshift distributions in each tomographic bin $\langle z \rangle_i$. The lower panels in each figure show the two dimensional posterior constraints on these parameters. Dashed grey lines correspond to mean values of the fiducial redshift distribution in each tomographic bin, and in the $S_8$ panel to the values inferred from a chain run without marginalisation over redshift nuisance parameters.
\subsubsection{Gaussian distributions for \deltaz}
\label{subsection:simulation-gaussian}
\begin{figure*}
    \centering
    \includegraphics[width=0.95\textwidth]{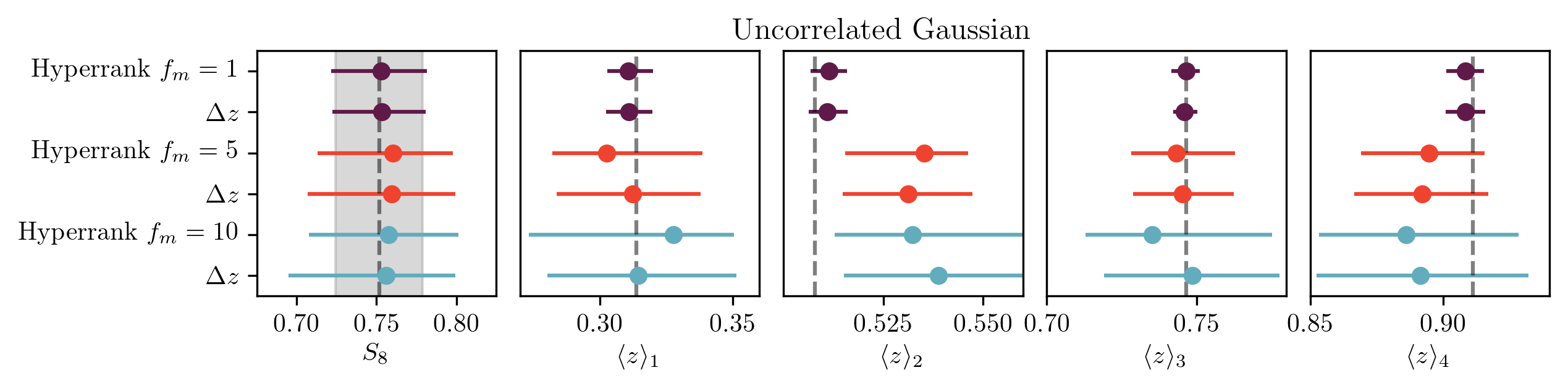}\\
    \includegraphics[width=0.33\textwidth]{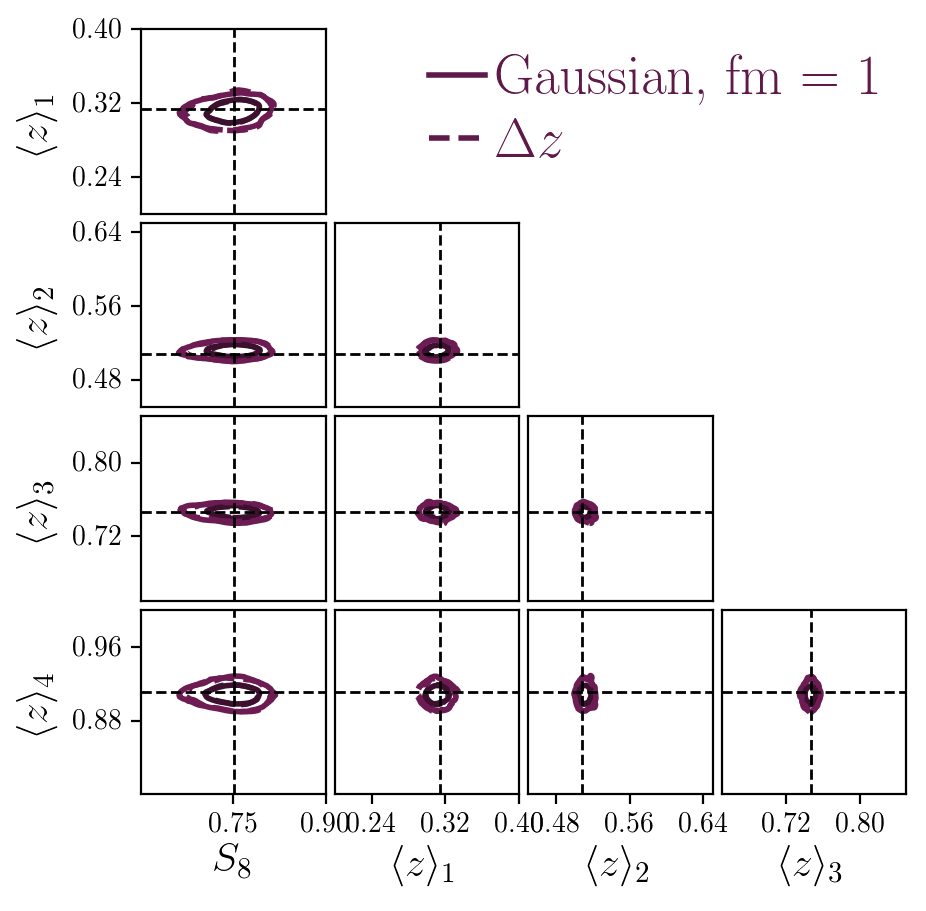}
    \includegraphics[width=0.33\textwidth]{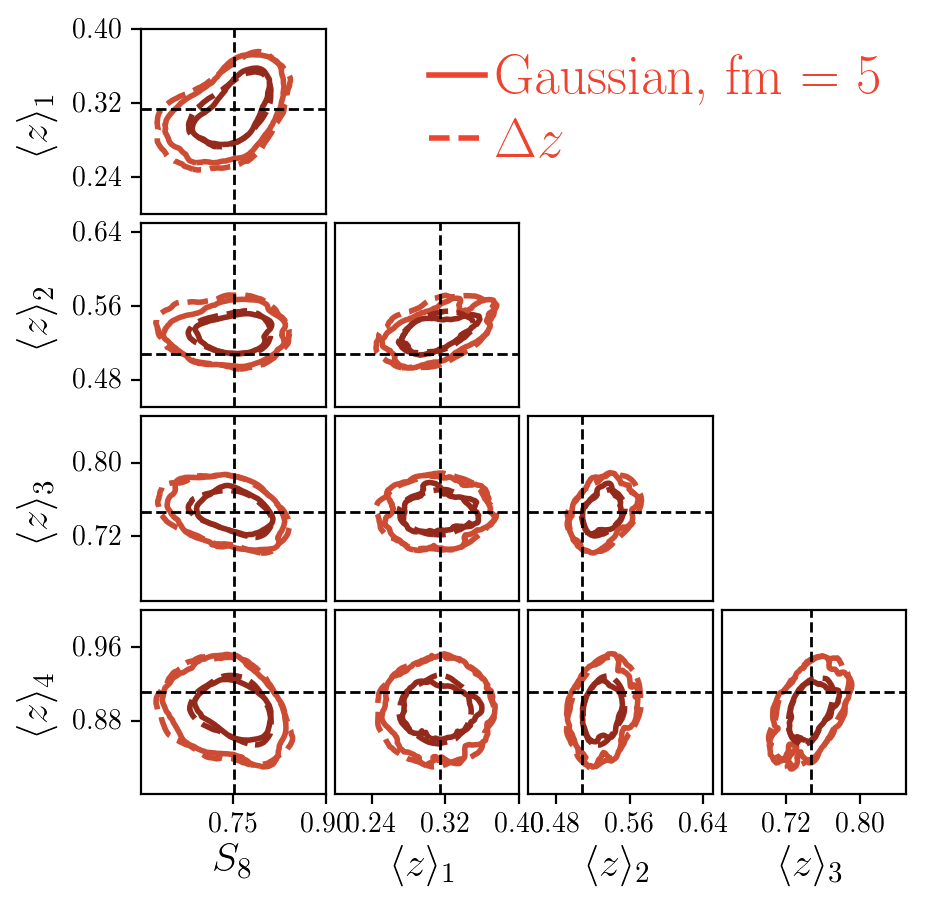}
    \includegraphics[width=0.33\textwidth]{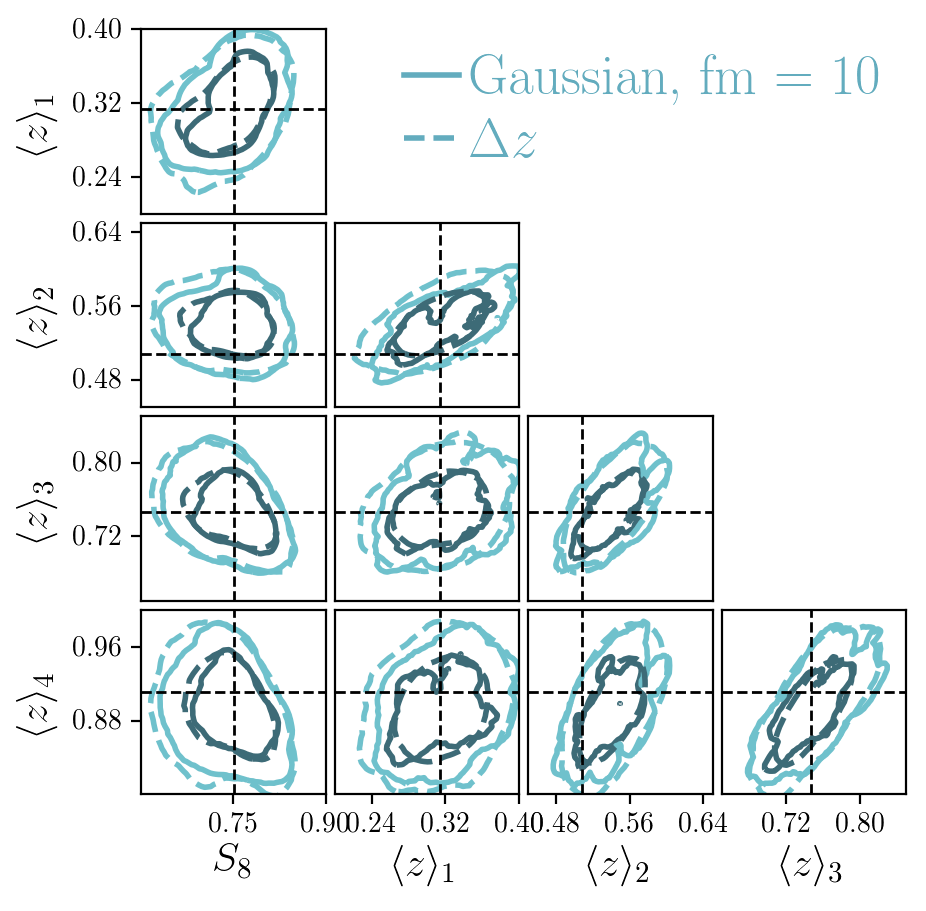}
    \caption{\emph{Upper}: Marginalised one-$\sigma$ confidence regions for the $S_8$, and $\langle z \rangle_i$ parameters for the uncorrelated Gaussian model of \nz uncertainty described in \cref{subsection:simulation-gaussian}. Different $f_m$ values refer to different overall amplitudes of uncertainty, and for each value of $f_m$ we show both the posterior from chains using \hyperrank and the \deltaz marginalisation schemes. \emph{Lower}: Two dimensional posteriors on the same parameters. Dashed grey lines correspond to mean values of the fiducial redshift distribution in each tomographic bin, and in the $S_8$ panel the $1\sigma$ region inferred from a chain run without marginalisation over redshift nuisance parameters is also shown.
    }
    \label{fig:results-gaussian}
\end{figure*}
We begin with the simple error model in which the \deltaz approach used in other analyses and described above is the correct one. Within each tomographic bin we draw values of \deltaz from a Gaussian distribution with width \sigmadeltaz. Realisations for \nz are then generated by shifting the fiducial \nzfid along the redshift axis by the drawn \deltaz. In order to assess performance and convergence we test this for several different levels of uncertainty, with the \sigmadeltaz being modified by a multiplicative factor $f_m$. For our fiducial \sigmadeltaz we use the values appropriate for DES-Y3 provided by \buzzard (see right panel of \cref{fig:cosmos_buzzard_realisations}). We then use values of $f_m = \lbrace 1, 5, 10 \rbrace$. We then run the full cosmological parameter estimation pipeline on the simulated data vector using these redshift distributions, once marginalising over the uncertainties using the Gaussian \deltaz method and once using the \hyperrank method on the set of realisations.

The results of this test are shown in \cref{fig:results-gaussian}. In the upper panel we show the one dimensional error bars recovered on $S_8 \equiv \sigma_8\left( \Omega_m/0.3\right)^{0.5}$ and the means of the redshift distributions in the four tomographic bins $\langle z \rangle_i$. Dashed grey lines correspond to mean values of the fiducial redshift distribution in each tomographic bin, and in the $S_8$ panel the $1\sigma$ region inferred from a chain run without marginalisation over redshift nuisance parameters is also shown. In the lower panel we also show the two dimensional posteriors for combinations of these parameters. We see that \hyperrank gives posteriors consistent with those obtained using the standard \deltaz marginalisation approach.
While at first glance this is a trivial example, it shows that the method is, at the very least, able to recover the same effects of redshift uncertainty when samples describe the same type of uncertainty we typically describe by means of a \deltaz nuisance parameter.

\subsubsection{Non-Gaussian distributions for $\Delta z$}
\label{subsection:simulation-nongaussian}
\begin{figure*}
    \centering
    \includegraphics[width=0.95\textwidth]{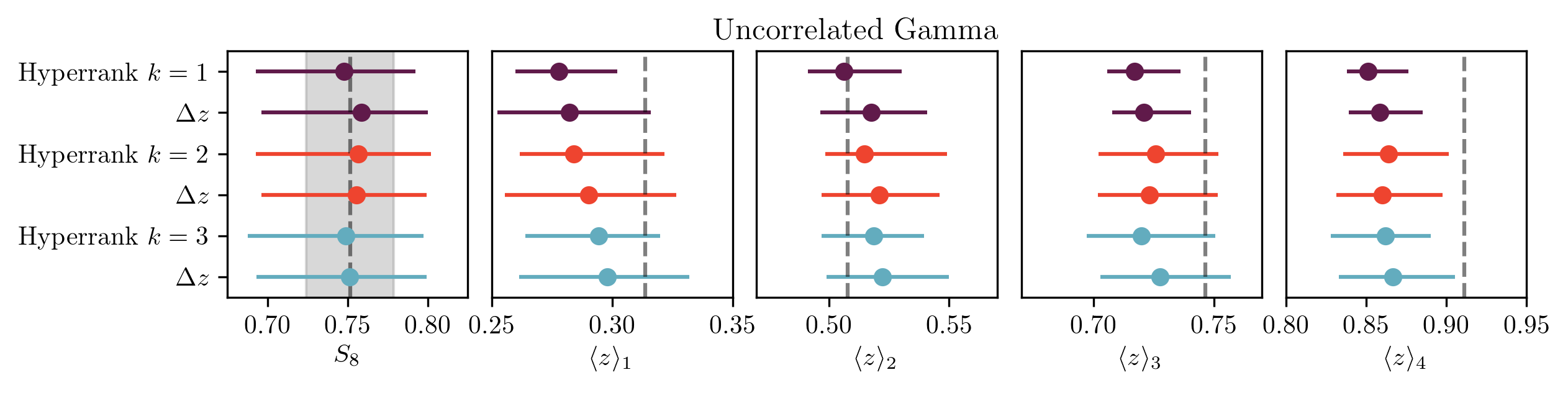}\\
    \includegraphics[width=0.33\textwidth]{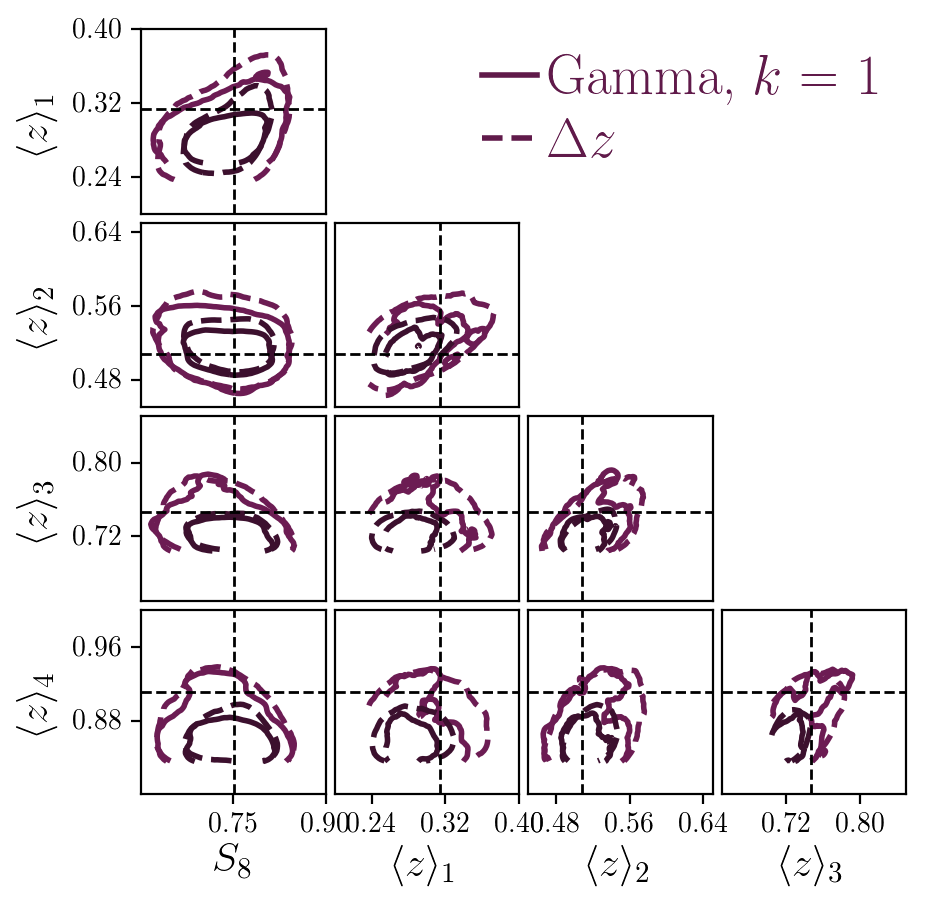}
    \includegraphics[width=0.33\textwidth]{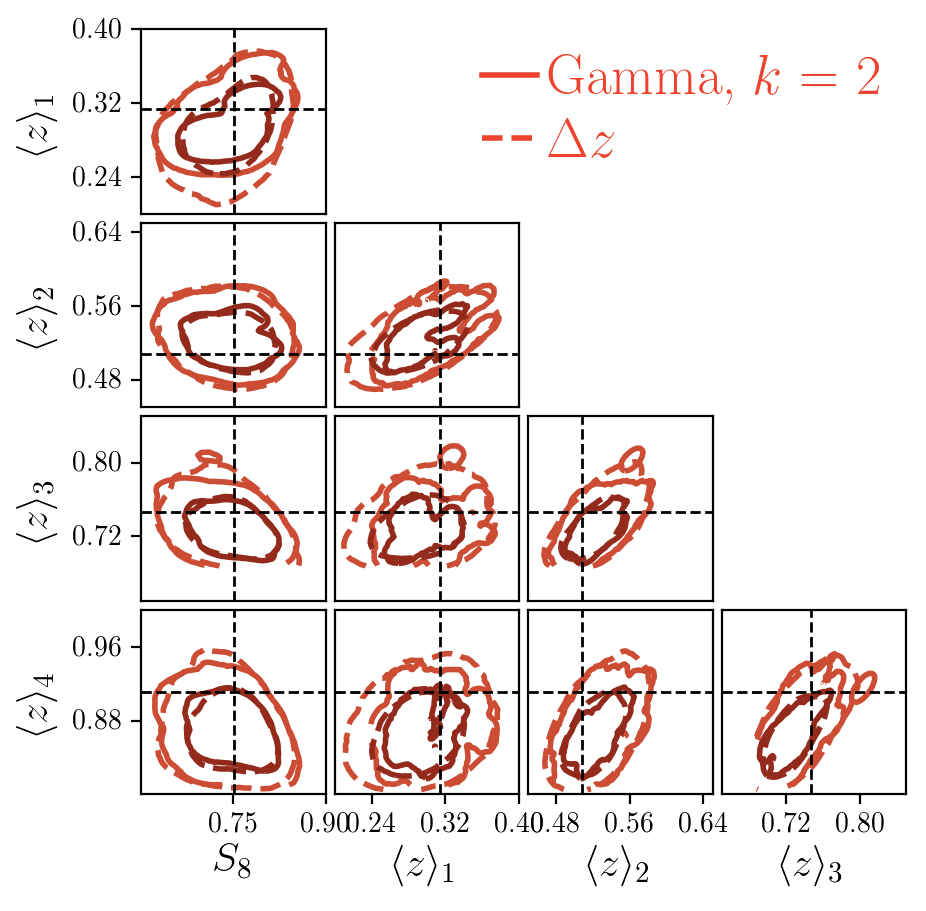}
    \includegraphics[width=0.33\textwidth]{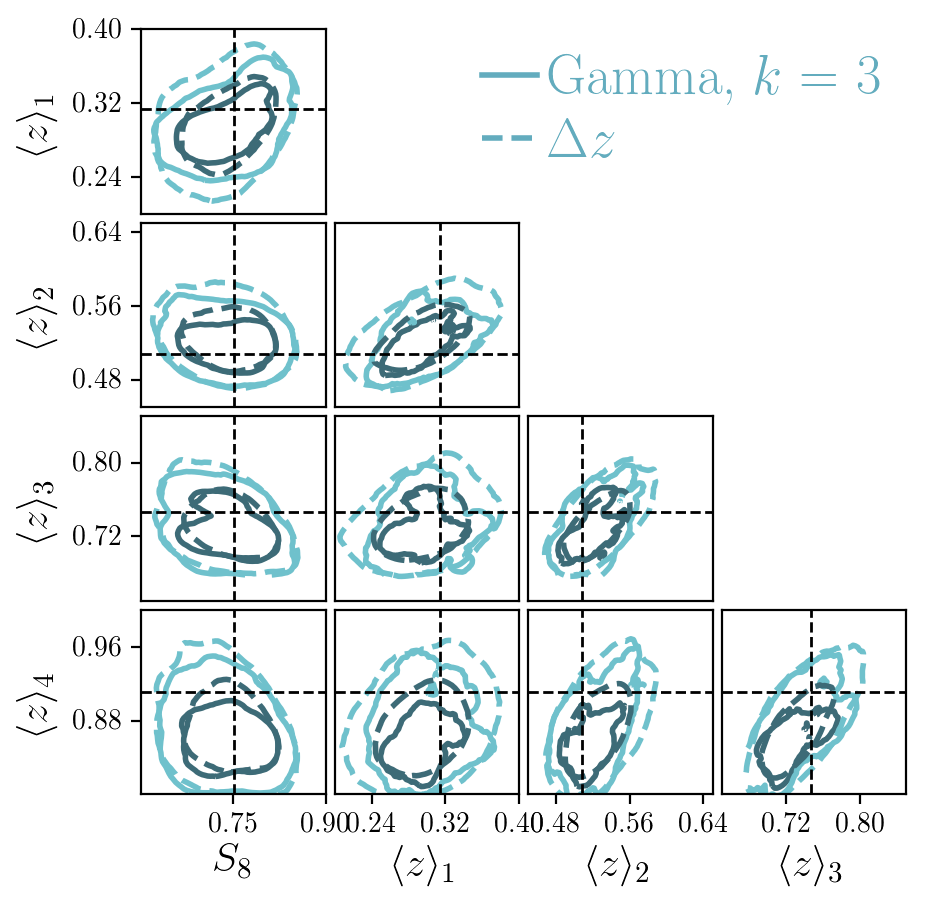}
    \caption{\emph{Upper}: Marginalised one-$\sigma$ confidence regions for the $S_8$, and $\langle z \rangle_i$ parameters for the uncorrelated Gamma distribution model of \nz uncertainty described in \cref{subsection:simulation-nongaussian}. Different $k$ values refer to different amounts of skewness in the uncertainty distributions and for each value of $k$ we show both the posterior from chains using \hyperrank and the \deltaz marginalisation schemes. \emph{Lower}: Two dimensional posteriors on the same parameters. Dashed grey lines correspond to mean values of the fiducial redshift distribution in each tomographic bin, and in the $S_8$ panel the $1\sigma$ region inferred from a chain run without marginalisation over redshift nuisance parameters is also shown.
    }
    \label{fig:results_nongaussian}
\end{figure*}
Modelling the distribution of \deltaz for each tomographic redshift bin as a Gaussian is a simple model choice which may not be an adequate representation of the true range and correlation structure of the \deltaz nuisance parameters, potentially resulting in a biased posterior and under/over-estimated uncertainties. 
In the right panel of \cref{fig:cosmos_buzzard_realisations} we show histograms of the \deltaz between the fiducial \nz and the 500 realisations generated using the full uncertainty model. These show appreciable non-Gaussianity, with skews and heavy tails which can be accentuated by the hard boundary at $z=0$ for all distributions, especially tomographic bins at low redshift.
We investigate the impact of the non-Gaussianity in the distribution of \deltaz by sampling \deltaz values from a highly skewed Gamma distribution:
\begin{equation}
    \label{eq:gamma_distribution}
    f(\deltaz; k, \theta) = \frac{\deltaz^{k-1} e^{-\deltaz/\theta}}{\theta^{k} \Gamma(k)}
\end{equation}
(where $\Gamma(k)$ is the integral Gamma function evaluated at $k$) to shift our fiducial distribution \nzfid. 
We use scale parameters $\theta$ such that the \sigmadeltaz for each tomographic bin is equal to that of the prior with the largest uncertainty in section \ref{subsection:simulation-gaussian} ($f_m = 10$, $\sigmadeltaz \sim 0.05$). We fix the shape parameter $k$ of the Gamma distribution to a set of values $k = 1,2,3$ to ensure the distribution of mean shifts of all tomographic bins have a positive skewness with a long tail to high values, and to explore the effect of different degrees of non-Gaussianity. The distribution of values is then centred so that the mean shift value is equal to zero, which generates a set of Gamma distributed \deltaz with the same variance and mean to that of the $f_m = 10$ prior, but with a skewness that cannot be captured by the use of a standard Gaussian prior. We then again run two chains, one marginalising over redshift uncertainty using the Gamma function \deltaz model, and one using \hyperrank on the generated realisations.

The result of these chains is shown in \cref{fig:results_nongaussian}. The differences on the $S_8$ parameter remain comparable to the typical dispersion seen for this number of distributions.
As in \cref{fig:results-gaussian} and \cref{fig:results_corrgaussian}, in the case of the $\langle z \rangle$ sampled values, small differences appear between \hyperrank and \deltaz chains, but they are distributed very similarly as seen in the lower panels of \cref{fig:results_nongaussian}, deviating in the same way from the reference values of the \nzfid distribution and no marginalisation run.
One aspect of the way \deltaz values are reported by \cosmosis can be responsible for the differences, as the $\langle z \rangle$ values shown here are the sampled \deltaz plus the means of the fiducial distribution \nzfid.
Because of the cut imposed at $z = 0$, this can result in a slightly inaccurate mean redshift value being computed here.

\subsubsection{Correlations between tomographic bins}
\label{subsection:simulation-tomographicbins}
\begin{figure*}
    \centering
    \includegraphics[width=0.95\textwidth]{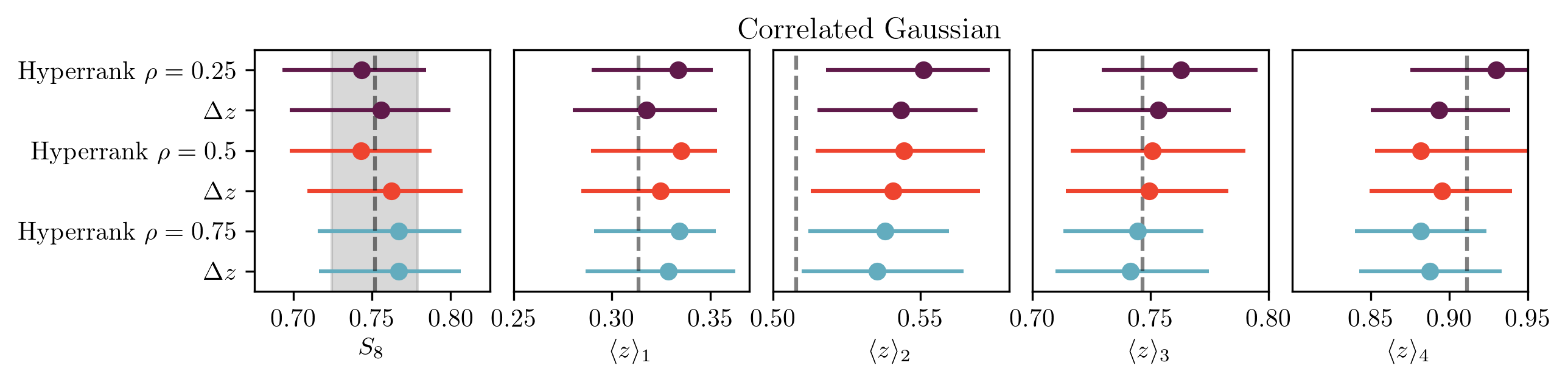}\\
    \includegraphics[width=0.33\textwidth]{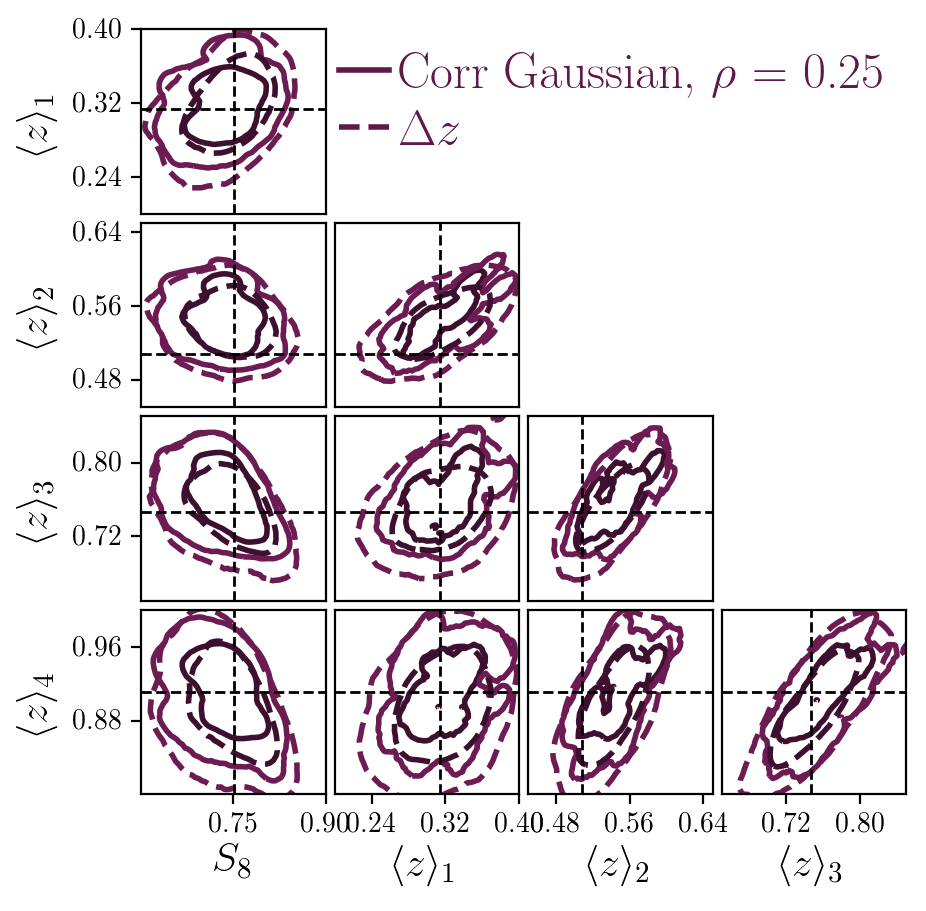}
    \includegraphics[width=0.33\textwidth]{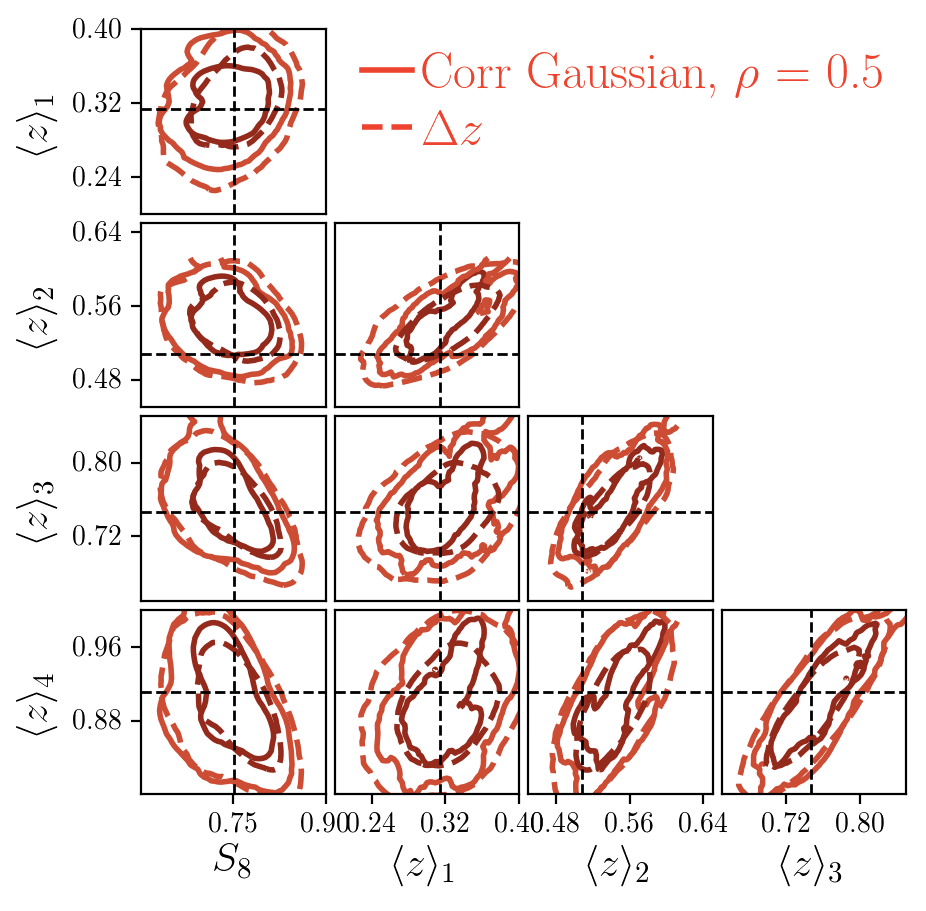}
    \includegraphics[width=0.33\textwidth]{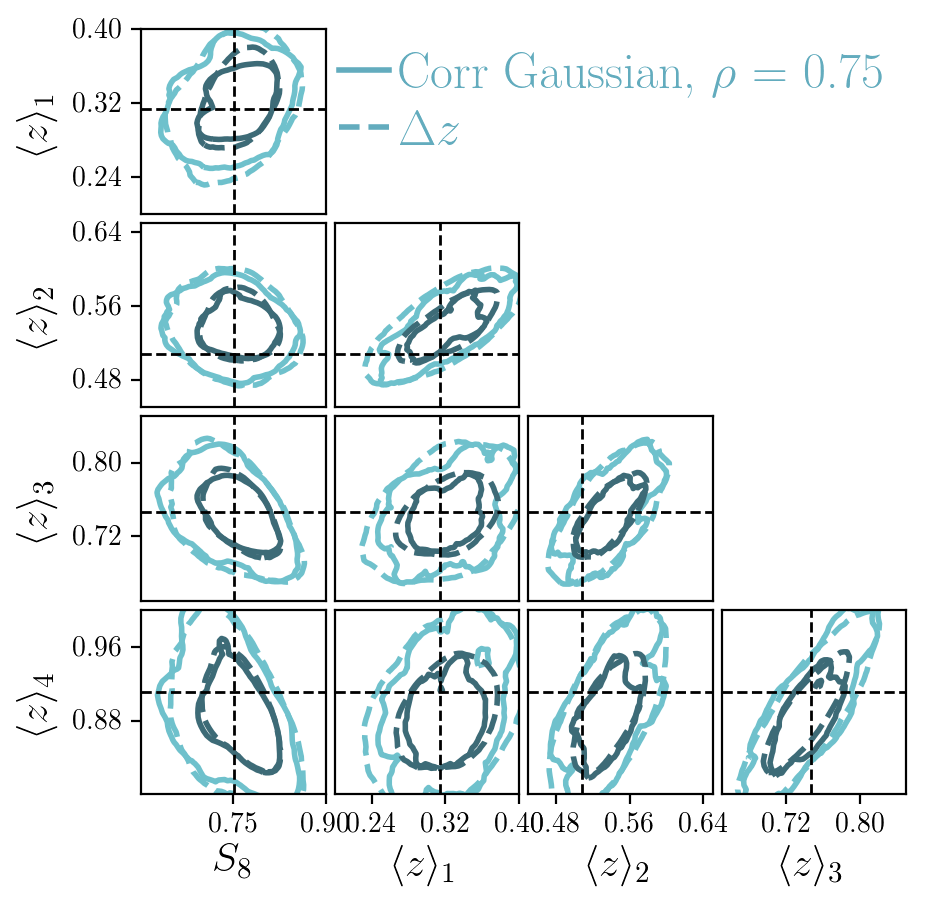}
    \caption{\emph{Upper}: Marginalised one-$\sigma$ confidence regions for the $S_8$, and $\langle z \rangle_i$ parameters for the correlated Gaussian distribution model of \nz uncertainty described in \cref{subsection:simulation-tomographicbins}. Different $\rho$ values refer to different amounts of correlation between tomographic bins in the uncertainty distributions and for each value of $\rho$ we show both the posterior from chains using \hyperrank and the \deltaz marginalisation schemes. \emph{Lower}: Two dimensional posteriors on the same parameters. Dashed grey lines correspond to mean values of the fiducial redshift distribution in each tomographic bin, and in the $S_8$ panel the $1\sigma$ region inferred from a chain run without marginalisation over redshift nuisance parameters is also shown.
    }
    \label{fig:results_corrgaussian}
\end{figure*}
Another aspect of uncertainty the simplest \deltaz scheme  does not directly account for is the potential correlation between the uncertainty from different tomographic bins \citep[though see Appendix A of][in which the diagonal elements of the covariance matrix are inflated to account for potential off-diagonal elements]{2018MNRAS.478..592H}.
Since each tomographic bin is shifted independently, combinations of \deltaz values which would not be expected to appear in multiple realisations of the survey or \photoz analysis are equally sampled.
In addition to this, the use of a single fiducial shifted \nz blurs the potential effect of correlation at the histogram bin level.
Correlation can come from the binning of galaxies and from how the shapes of the distributions and their moments can change when galaxies are re-assigned to another histogram or tomographic bin in a different realisation of a photo-z analysis.
Depending on the nature of the colour-redshift degeneracy, correlation can also appear between non contiguous tomographic bins.

In this case, the standard \deltaz scheme can not be expected to preserve the effects of such correlations, as the set of $N_{\rm tomo}$ \deltaz nuisance parameters are sampled independently from their corresponding priors in the Monte Carlo chain.
By contrast, drawing a value of the \hyperrank parameter(s) in a chain jointly specifies the \nz to be used in all tomographic bins and preserves these correlations, which can potentially lead to tighter contours on the cosmological parameters since the space of \deltaz values is restricted to those allowed by the samples.
Depending on the sign of the correlation, this can also result on a shift of the contours if the \deltaz values favour a combination of high or low mean redshift only (positive correlation), instead of a combination of low and high mean redshift (negative correlation). To explore the potential effects of these correlations at the tomographic bin level on inferred cosmological parameters, we generate three sets of mean-shifted realisations of the fiducial \buzzard \nzfid by values of \deltaz sampled from a covariance matrix with increasing correlation between tomographic bin pairs (1,2) and (3,4). This is intended to be a simple model of leakage of galaxies between adjacent tomographic bins, with more complicated models for bin correlation also possible.
We generate the samples so their Pearson correlation coefficients take the values $\rho = \{0.25, 0.5, 0.75\}$, and employ the same coefficient for both bin pairs while leaving all other bin pairs uncorrelated ($\rho=0$).
To better visualise the effects of these correlations once again we use an amplified \sigmadeltaz prior to describe the diagonal of the covariance matrix, equal to the $f_m=10$ prior described in \cref{subsection:simulation-gaussian}. We again run two chains, one in which a correlated Gaussian \deltaz marginalisation is used by drawing values from a correlated prior with the same correlation matrix to that used to generate the proposal samples, and one in which a \hyperrank marginalisation is used.

The result of this test is shown in \cref{fig:results_corrgaussian}. We can again see that \hyperrank correctly recovers the uncertainty in \deltaz and $S_8$ represented by the two dimensional posteriors.

\subsubsection{Higher order modes of uncertainty}
\label{subsection:simulation-peculiarities}
\begin{figure*}
    \centering
    \includegraphics[width=0.95\textwidth]{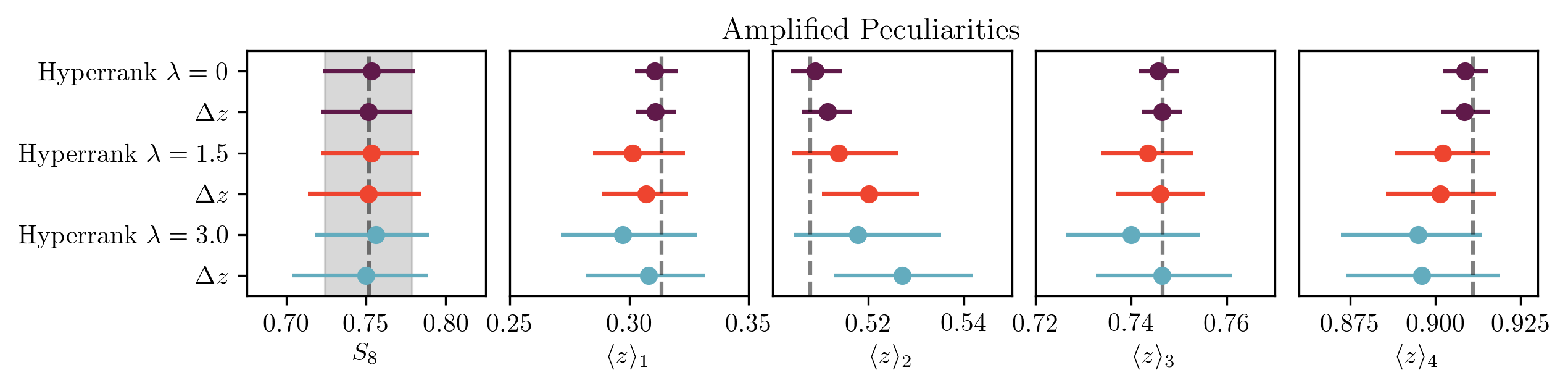}\\
    \includegraphics[width=0.33\textwidth]{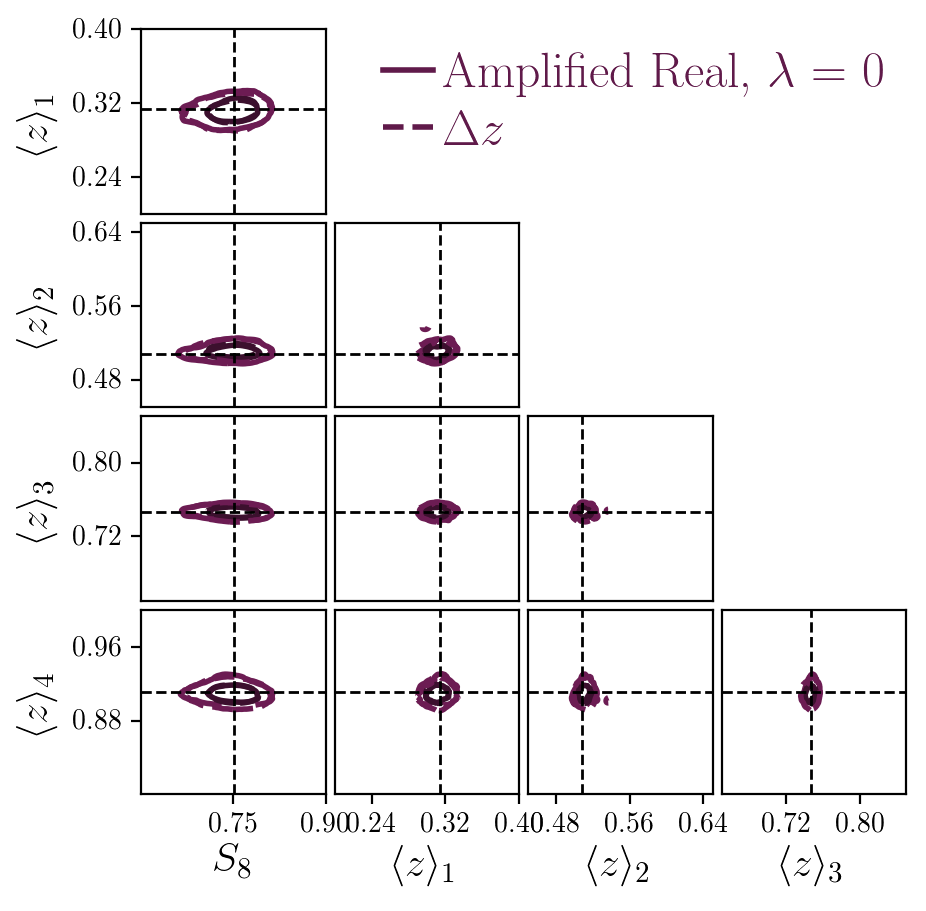}
    \includegraphics[width=0.33\textwidth]{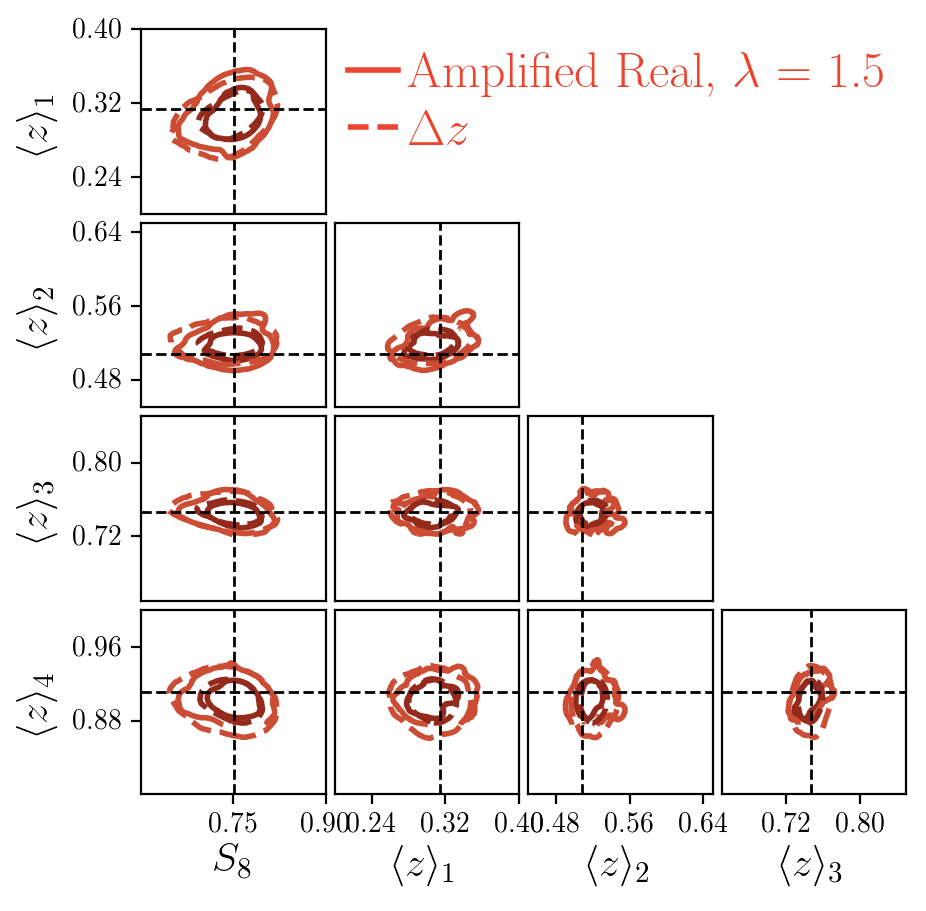}
    \includegraphics[width=0.33\textwidth]{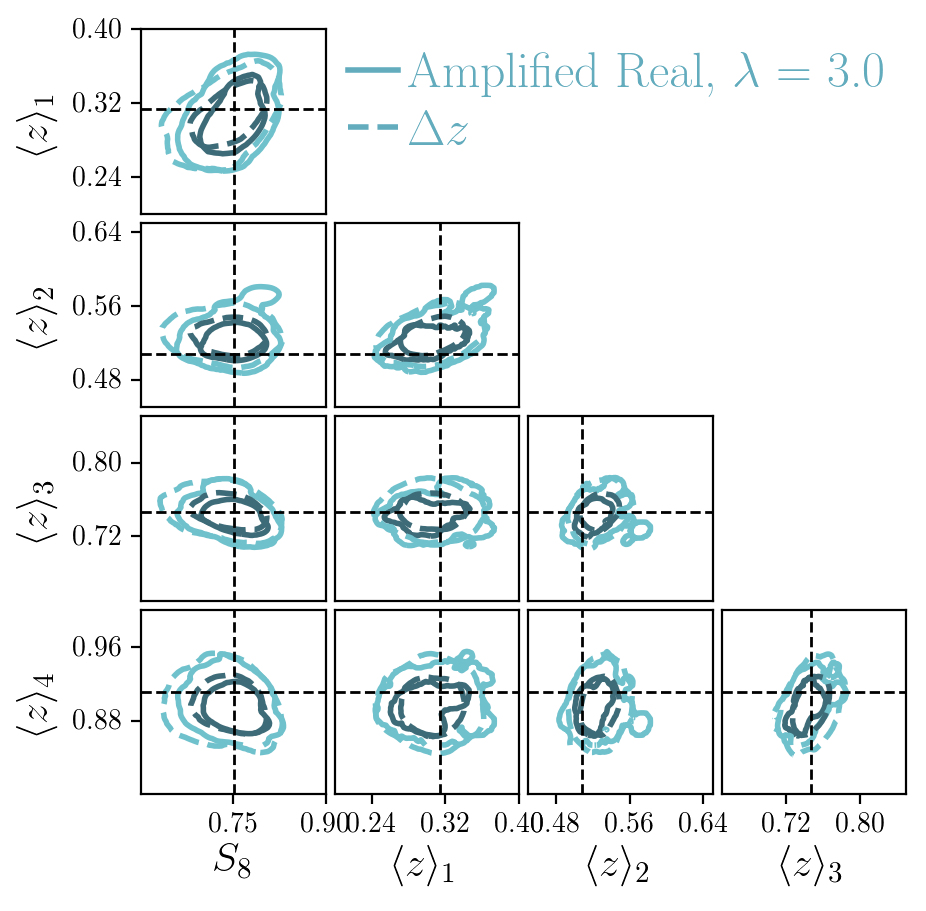}
    \caption{\emph{Upper}: Marginalised one-$\sigma$ confidence regions for the $S_8$, and $\langle z \rangle_i$ parameters for the amplified deviations model of \nz uncertainty described in \cref{subsection:simulation-peculiarities}. Different $\lambda$ values refer to different amplifications uncertainty in the \nz distributions and for each value of $\lambda$ we show both the posterior from chains using \hyperrank and the \deltaz marginalisation schemes. \emph{Lower}: Two dimensional posteriors on the same parameters. Dashed grey lines correspond to mean values of the fiducial redshift distribution in each tomographic bin, and in the $S_8$ panel the $1\sigma$ region inferred from a chain run without marginalisation over redshift nuisance parameters is also shown.
    }
    \label{fig:results_amplified}
\end{figure*}
The above tests show that \hyperrank is capable of correctly marginalising over redshift distribution uncertainties in cases where a correct and simple model for them is known. Finally, in this section, we use a set of realisations of \nz which represent a fully flexible model of the uncertainty in \nz, following the approach of \cite*{y3-sompz} and summarised in section \ref{sec:sompz}, as applied to the \buzzard simulation.

As above for the cases of different values of $f_m$, we apply a procedure to these realisations to artificially increase the level of uncertainty they represent. Starting from the set of 500 realisations, we amplify the difference between each of the $n(z_i)$ values and the value of the fiducial distribution, $n^f(z_i)$, such that $n'(z_i) = n(z_i) + \lambda \left[n(z_i) - n^f(z_i)\right]$.
Hence, we decide to use the typical dispersion values found for 500 realisations for that amplification as the reference to evaluate the contours obtained with \hyperrank.
For this test we generate three sets of distributions: one with no amplification, $\lambda = 0$; and two with amplified peculiarities, $\lambda = \{1.5,\, 3\}$.
While the average \nz obtained from the amplified realisations remains unaltered, this procedure can result in a slightly wider equivalent Gaussian prior \sigmadeltaz to those of the un-amplified realisations.
Thus, we also obtain the \sigmadeltaz values for each set of distributions and use them to compare \hyperrank to the standard \deltaz marginalisation.

The results from this test are shown in \cref{fig:results_amplified}. As can be seen, the \hyperrank and \deltaz chains again recover highly consistent contours on the $S_8$ and $\langle z \rangle$ parameters. For the $\lambda=0$ case the \deltaz posterior on $S_8$ is 2.6\% narrower than the \hyperrank one, for $\lambda=1.5$ the \deltaz posteriors are 16\% wider, and for $\lambda=3.0$ the \deltaz posterior is 18\% wider. This follows the idea that \hyperrank is capable of better modelling of these more complex uncertainties, but that in the $\lambda=0$ (DES-Y3-like) regime, \deltaz is an acceptable approximation.

\subsection{Sampling efficiency and ranking mode}
\label{subsection:simulation-efficiency}
As well as the correct exploration of the uncertainties, we also wish to see the effect of the \hyperrank procedure on the efficiency of mapping the posterior of cosmological and nuisance parameters.
For a randomly sampled set of distributions the likelihood is not a smooth function of the parameters being sampled (see \cref{fig:1d-logposterior}).
Therefore, the parameter space volume cannot be sampled consistently in higher likelihood regions since there is no correlation between the sampled nuisance parameter and cosmology posterior. Any proposal step in the Monte Carlo algorithm typically does not have the intended effect, since proposed jumps in the redshift nuisance parameters are now across a random, discontinuous likelihood.
This leads to the sampler requiring many more likelihood evaluations to find new samples of the posterior.
We define sampling efficiency $\eta$ as the number of replacements (samples of the posterior) made by \multinest over the total number of likelihood evaluations required for convergence, with higher $\eta$ representing better performance.

We test the different mapping schemes described in \cref{section:hyperrank} comparing 1D and 3D $\langle z \rangle$, 3D $\langle 1/\chi \rangle$ and a KL approach where the first K=3 components are used. We compare the sampling efficiency between them and against a naive sampling where realisations are chosen at random on each likelihood evaluation.
To reduce the effect of sampling noise due to the stochastic nature of the sampler, we repeat each run five times with different initial random seeds for the sampler.
\begin{figure}
    \centering
    \includegraphics[width=\columnwidth]{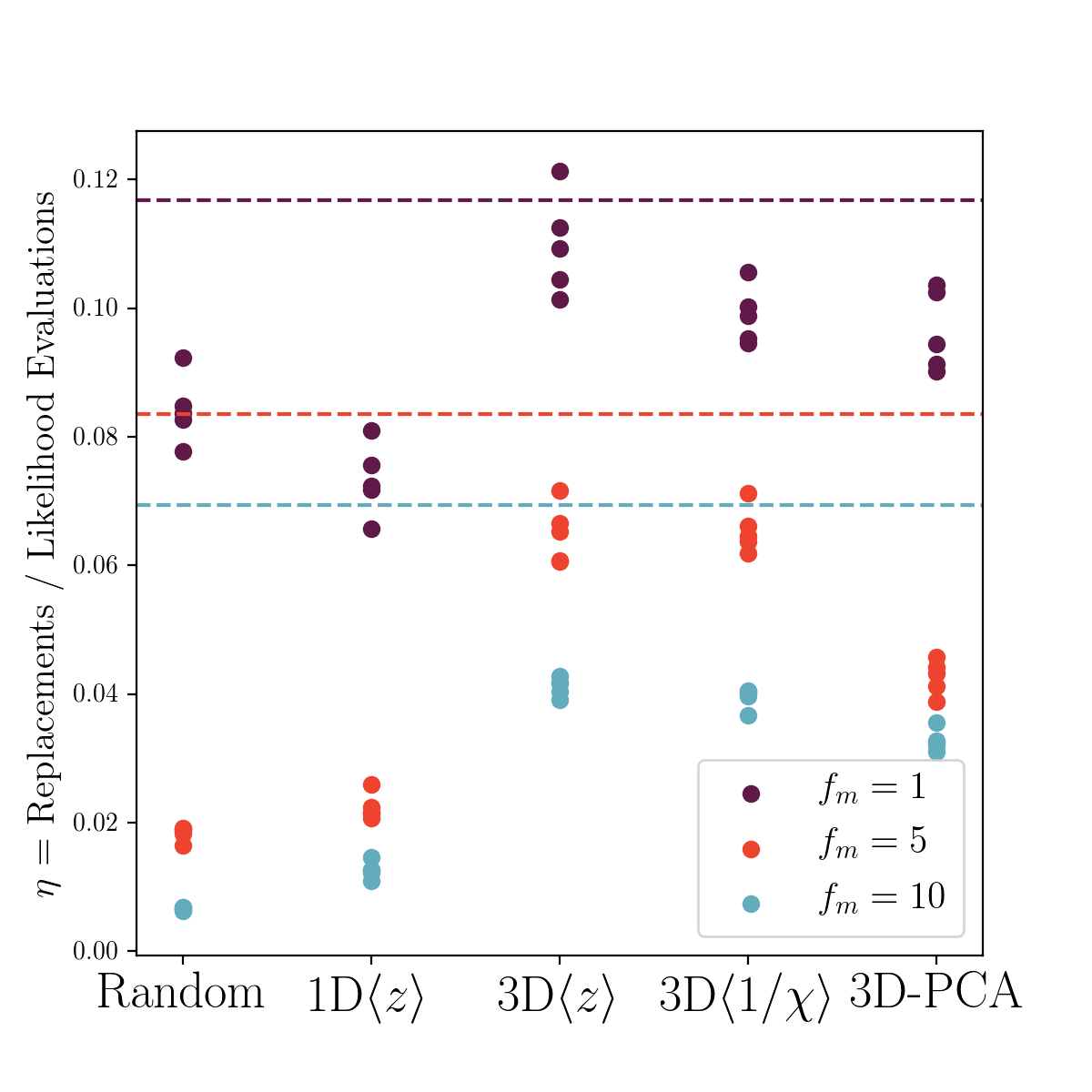}
    \caption{
    The sampling efficiency $\eta$ for the different mapping schemes described in \cref{section:hyperrank}, for three prior amplifications values $f_m$ ($f_m=1$ is the original \buzzard redshift distributions). The spread of points at each location shows the differences due to different initial random seeds.
    The rankings are ordered from left to right as a function of perceived complexity, with a random ranking being the most naive approach and a 3D KL corresponding to the most complex to implement.
    Horizontal dashed lines show the efficiencies obtained by marginalising the same equivalent uncertainty \sigmadeltaz using the \deltaz method, obtained after averaging 5 runs with each equivalent prior.}
    \label{fig:sampling_efficiency}
\end{figure}

\Cref{fig:sampling_efficiency} shows the sampling efficiencies $\eta$ at different $f_m$ values as a function of different choices for descriptive values $\mathbf{d}$, all compared to the average efficiency from five runs  obtained using the \deltaz approach (dashed horizontal lines). The different colours used represent this test for different values of the $f_m$ parameter.
In all cases it is clear that the more complex choices of $\mathbf{d}$ using multiple dimensions are more efficient at exploring the space of uncertainties, with 3D $\langle z \rangle$ and 3D $\langle 1/\chi \rangle$ showing better performance at all $f_m$ values.
This is expected since the addition of more dimensions helps breaking the degeneracy of the posterior values present when a single parameter is used and all the information of the \nz realisations is compressed into a single value.

The KL approach, also tested in three dimensions, provides an improvement over random and 1D sampling, but does not reach the same levels of efficiency for methods of equal dimensionality.
with respect to a reference data vector obtained at a fixed cosmology, and the relative importance of each \nz element can change as the sampler moves in cosmology space.

Perhaps one surprising result occurs when comparing the random approach against 1D $\langle z \rangle$ in the un-amplified case ($f_m=1$), in which the former appears $\sim10\%$ more efficient.
We believe this is caused by the relatively small contribution of \nz uncertainty to the posterior in the $f_m=1$ regime, as all realisations have very similar mean values across all tomographic bins.
This can lead to a very small change of smoothness of the posterior at a fixed cosmology when moving from a random ordering to a 1D ordering, resulting in similar efficiencies.
While we do not show the effect of additional dimensions for a similar type of descriptive value $\mathbf{d}$ (i.e. 4D $\langle z \rangle$), some test runs suggest their efficiency is not noticeably better than a 3D approach, at the expense of noisier posteriors on the $\mathcal{H}$ parameters.
\response{This is likely caused by the larger discontinuities in the posterior surface as a function of $\mathcal{H}$, which is a consequence of the lower resolution of the multidimensional grid in higher $N_d$ (as discussed in \cref{subsection:hyperrank-multid})}.

Based on these results we consider a 3D approach an appropriate default configuration, with a preference for $\langle z \rangle$ since its computation does not involve the use of a fiducial cosmology, unlike $\langle 1/\chi \rangle$ (which requires calculation of $\chi(z)$ for \cref{eq:g}, done here and typically elsewhere at a fixed cosmology). When considering which three of the four tomographic bins to choose to use as \hyperrank dimensions, we recommend that tomographic bins should be ordered in terms of the variance in the descriptive value (e.g. the spread of different mean redshifts across different realisations), with the tomographic bin with lowest variance in the descriptive value not used as a \hyperrank dimension.

\begin{figure}
    \centering
    \includegraphics[width=\columnwidth]{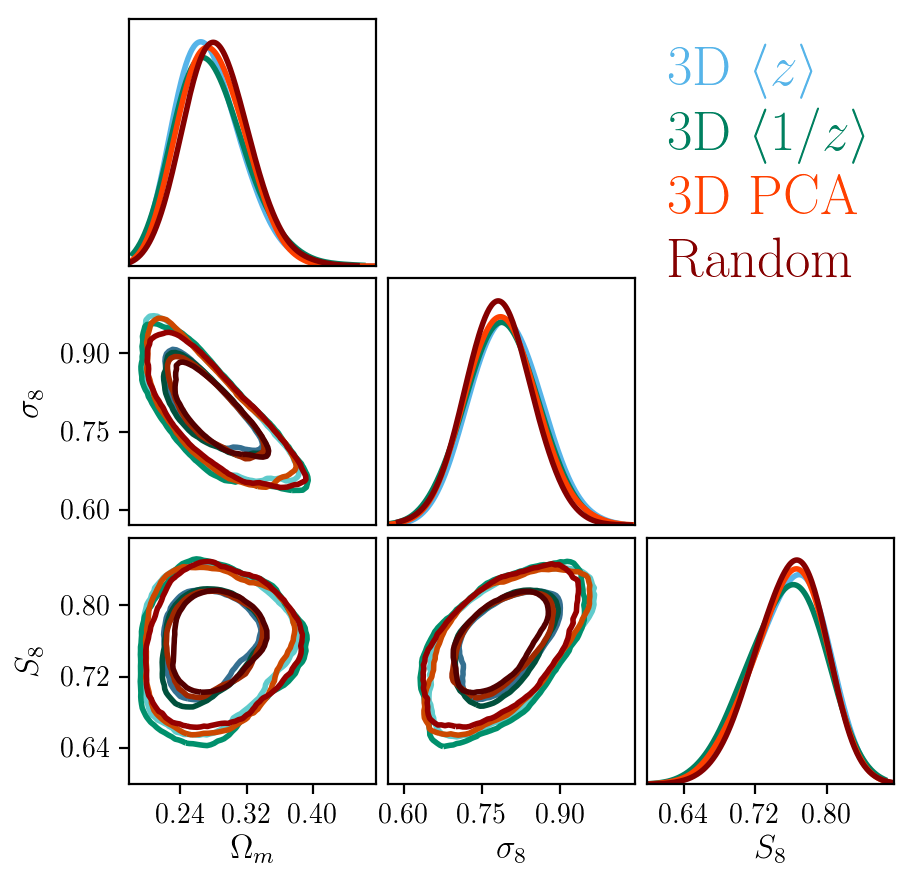}
    \caption{Recovered cosmological parameters for four different ranking approaches described in the text: Three dimensional rankings by mean redshift and inverse comoving distance of tomographic bins, three dimensional ranking by principal components of the data vector, and random sampling of realisations.
    The contours shown correspond to the case where realisations sampled with hyperrank describe an uncorrelated Gaussian distribution with an amplification factor of the uncertainty $f_m = 5$.
    For $f_m = 1, 10$ the contours are also very similar between the different ranking schemes.}
    \label{fig:ranking_contours}
\end{figure}

In \cref{fig:ranking_contours} we show posterior contours recovered for each of the different ranking schemes, including the `Random' scheme in which no ranking is performed. The consistency of these contours confirms that the \hyperrank procedure does not affect the cosmology recovered whilst improving the efficiency of a chain with respect to un-ranked, random sampling of \nz realisations.

\subsubsection{Convergence}
In \hyperrank, discrete samples from the posterior over the subset of redshift nuisance parameters are generated outside of the main chain used to sample over the cosmological and other nuisance parameters. This means a limited and discrete set of values of the nuisance parameters are available to the main sampling, as opposed to the continuous range of parameters within a specified prior which would be available otherwise. 
\response{Whilst it could be possible to form an interpolation between closely-ranked samples to boost their number, such interpolated samples would no longer carry the intended property of being true samples from the posterior for $n(z)$, with the interpolation scheme effectively becoming part of the model and correlation structure for histogram bin heights, with a set of hidden hyperparameters. It is also unclear how such an interpolation could include the correlations across tomographic bins which we have found to be an important describing factor of the samples.}

\response{Without interpolation, there} will be a transition from the regime in which there are too few realisations of \nz available to effectively explore the redshift distribution uncertainty, and the limit where infinitely many realisations would be available, corresponding to the continuous case. Here we investigate the convergence of \hyperrank marginalisation with respect to the number of \nz samples generated, for the case of our DES-Y3 simulated data set.

\begin{figure}
    \centering
    \includegraphics[width=\columnwidth]{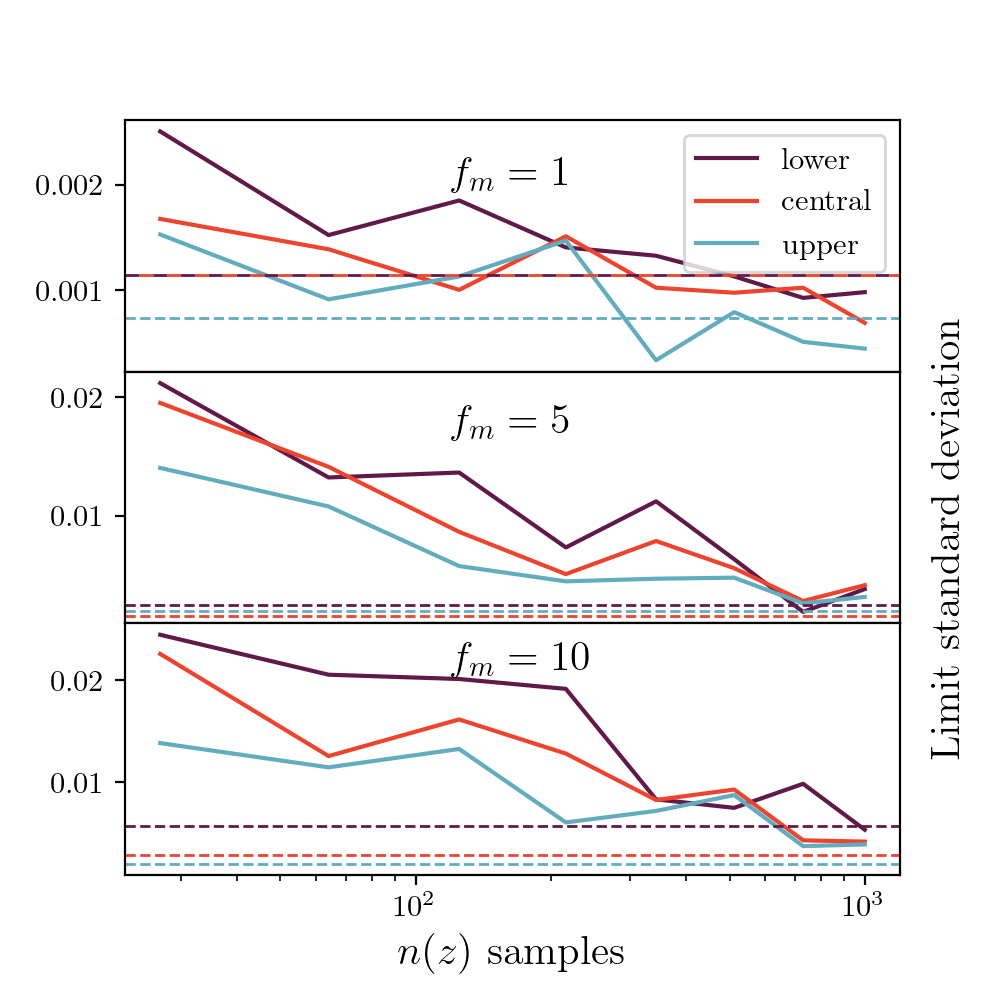}
    \caption{Standard deviation of the lower (purple), central (red), and upper (cyan) values for the $S_8$ parameter obtained using \hyperrank for 5 realisations of the ensemble of \nz samples, as a function of the total number of distributions to form the ensemble.
    From top to bottom, the equivalent \sigmadeltaz width is amplified by a progressively larger number, $f_m$, with respect to the original distributions of \buzzard samples.
    Horizontal dashed lines indicate the typical standard deviation for runs using the traditional \deltaz marginalisation approach.}
    \label{fig:number_samples_std}
\end{figure}

We first generate several sets of distributions where each realisation is a shifted version of a fiducial \nzfid, and the shifts are drawn from a Gaussian prior, following a similar approach to the $\deltaz$ method described in \Cref{section:y3-discrete-redshifts}.
We generate 8 sets of redshift distributions, each containing $3^3, 4^3, 5^3, 6^3, 7^3, 8^3, 9^3, 10^3$ realisations which are then ranked using the 3 dimensional default configuration described at the end of \cref{subsection:hyperrank-multid}.

Since we expect the approximate minimum number of realisations required for this convergence to depend on the level of uncertainty in the \nz, we generate two additional sets of proposal distributions by multiplying the \sigmadeltaz obtained above, by a factor $f_m = {5, 10}$. We then repeat the generation of proposal realisations with five different random seeds for each of the three $f_m$ values, and for each of the 8 sets of realisations containing different number of proposals.
By comparing the standard deviation on the central, lower and upper confidence values for $S_8$ as a function of the number of realisations, we can find an approximate minimum number of realisations required for the standard deviation of error bars from \hyperrank to converge to that obtained using the \deltaz approach (which is formally correct for this set of realisations).
In \cref{fig:number_samples_std} we observe that for all three levels of uncertainty, described by the amplification factor $f_m$, 1000 realisations yield standard deviation of the error bars obtained using \hyperrank comparable to the ones using the \deltaz approach.

\section{Application to DES Year 3}
\label{section:y3}
\begin{figure*}
    \centering
    \includegraphics[width=\columnwidth]{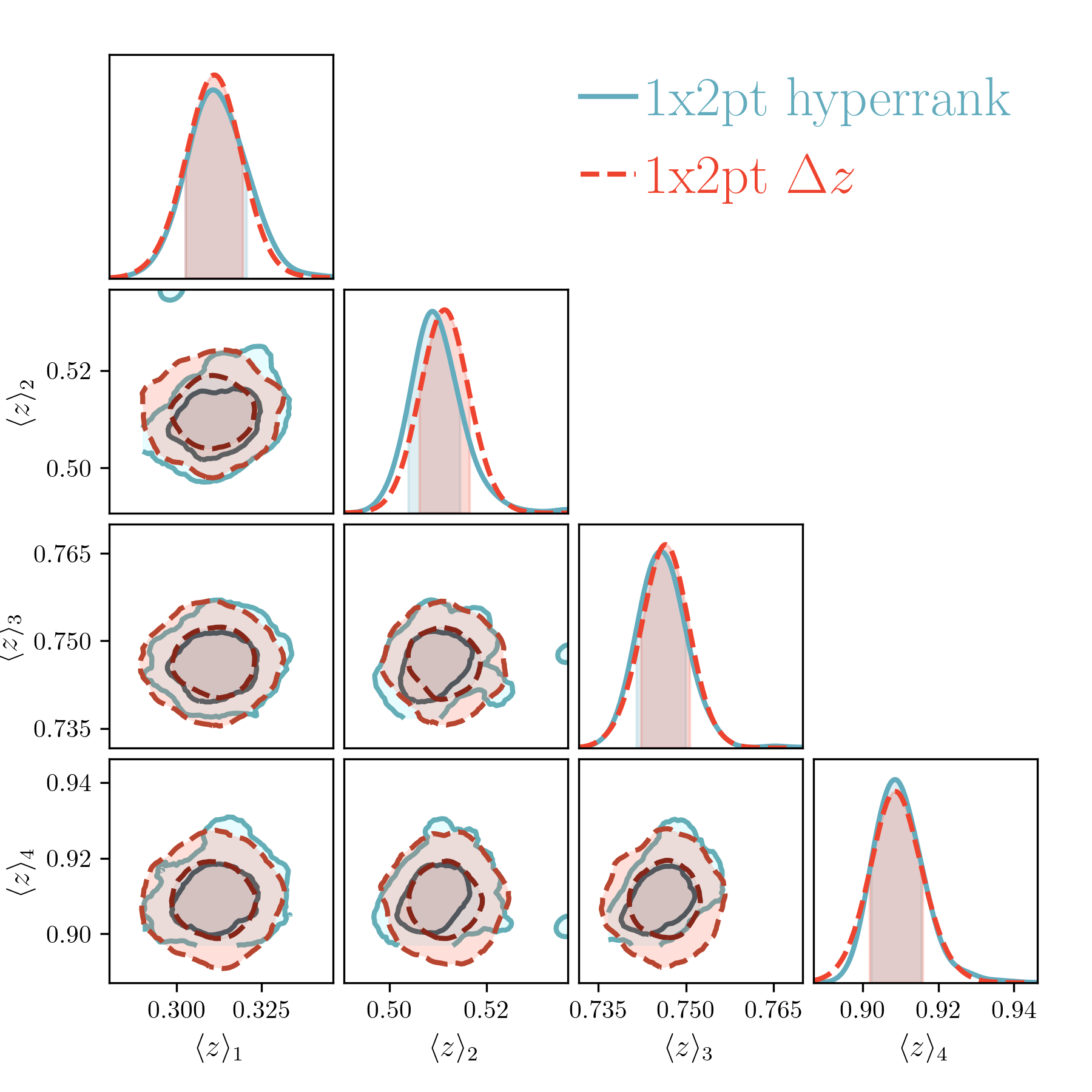}
    \includegraphics[width=\columnwidth]{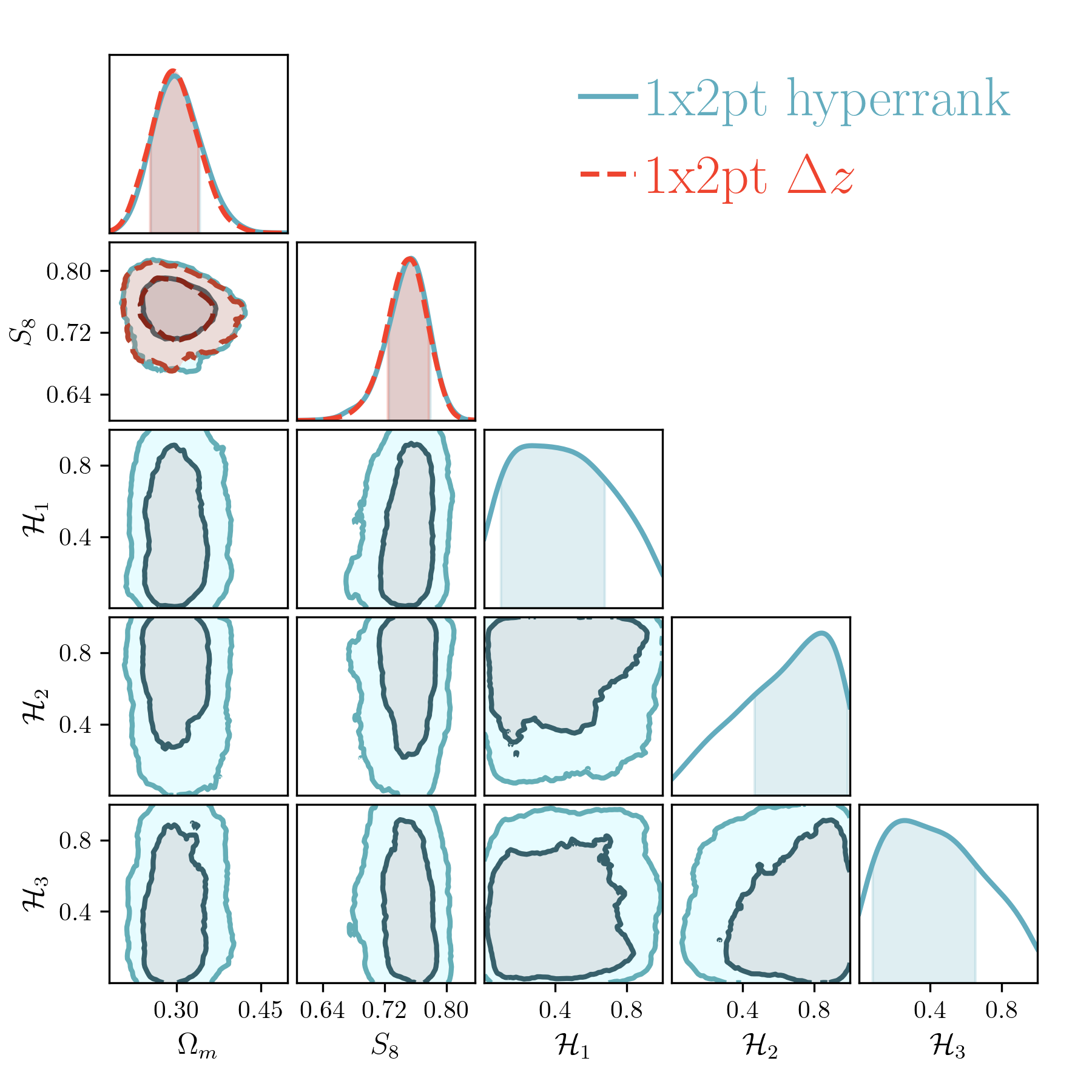}
    \caption{
    Correlation between redshift distribution uncertainty nuisance parameters in the \buzzard simulated DES-Y3 analysis, comparing the standard \deltaz approach (red) with the \hyperrank approach presented in this work (blue). \emph{Left} shows the recovered posteriors on mean redshifts of redshift distributions within the tomographic bins considered. \emph{Right} shows the recovered cosmological parameters for both approaches, and the \hyperrank ranking parameters. Both show good agreement between the two approaches for the modelled uncertainty expected in DES-Y3.
    }
    \label{fig:posterior-y3-hyperrank}
\end{figure*}

Based on the above tests, we derive an appropriate configuration for using \hyperrank on DES-Y3 (or similar) data:
\begin{itemize}
    \item $\langle z \rangle$ ranking
    \item Three \hyperrank parameter dimensions
    \item Ranking according to tomographic bins 1, 2 and 4
    \item At least $10^3$ \nz samples available to \hyperrank
\end{itemize}
 We then run a full shear-only cosmology chain on the \buzzard simulation of the DES-Y3 data set, with model parameterisations and priors as discussed in the main cosmology papers (\citealt{y3-cosmicshear1}; \citealt*{y3-cosmicshear2}), and with 1000 realisations of possible \nz generated using the full procedure of \cite*{y3-sompz}. This as closely as possible mimics the experimental data and setup of the DES-Y3 analysis. We also run a chain with this setup, but with the \hyperrank marginalisation of redshift uncertainties replaced by the \deltaz approach. This results of these two chains are shown in \cref{fig:posterior-y3-hyperrank}. The left panel shows the posteriors on mean redshift within the four tomographic bins, produced directly by the \deltaz analysis and by taking the posterior weighted means within the \hyperrank analysis. Good consistency can be seen in the space of mean redshifts explored. The right panel of \cref{fig:posterior-y3-hyperrank} shows posteriors on the cosmological parameters, and the \hyperrank parameters for each tomographic bin. For the cosmological parameters we also show the recovered posterior from the \deltaz analysis, finding highly consistent results between the two approaches. This suggests that for the uncertainties which are modelled as part of the DES-Y3 analysis, the \deltaz approach is adequate to fully explore their effect on cosmic shear cosmological parameters. The \deltaz approach is hence adopted as fiducial in \cite{y3-cosmicshear1} and \cite*{y3-cosmicshear2} and subsequent DES-Y3 analyses, with the validation test between \hyperrank and \deltaz shown here on the \buzzard simulation repeated for the real data vector in section E.1. of \cite{y3-cosmicshear1}.

We also show the recovered posteriors on \hyperrank ranking parameters, showing that different subspaces of the ranked \nz realisations are indeed favoured in a systematic way, indicating the cosmological data are in turn helping constrain the space of plausible redshift distributions.

\section{Conclusions}
 \label{section:conclusions}
We have presented \hyperrank, a new approach to marginalise over redshift distribution uncertainties in weak lensing and galaxy clustering experiments by ranking and mapping a set of proposal redshift distributions to a set of continuous hyper-parameter which are then sampled in the Monte Carlo chain.

To test the accuracy of the method we generated a series of \nz ensembles to describe different types of uncertainty, and compared the obtained $S_8$ error estimates and sampled uncertainty to those obtained by only marginalising over a shift \deltaz along the redshift direction for each tomographic bin. 

We showed that this approach provides equivalent results to the ones obtained marginalising over \deltaz, when the realisations of the ensemble are obtained by shifting the tomographic bins of a fiducial distribution by a set of values drawn from the same prior used to describe the \deltaz uncertainty.

We generated additional ensembles to represent types of uncertainty which cannot be fully characterised by a set of uncorrelated Gaussian shifts \deltaz, and if unaccounted for, can lead to an incorrect estimation of the marginalised cosmological parameters posteriors.
These included samples with \deltaz shifts drawn from non-Gaussian distributions, drawn from highly correlated multivariate Gaussian distributions and from a set of realistic distributions with amplified peculiarities, based on the estimates obtained from the \sompz scheme on the \buzzard simulations.
In all cases, \hyperrank correctly explores the uncertainty described by the input distribution ensemble, providing posteriors on the cosmological $S_8$ and redshift tomographic bin means $\langle z \rangle$ which are highly consistent with those from the estimates obtained using marginalisation with \deltaz nuisance parameters which are distributed according to the input model (and hence are the correct model for the uncertainty).

A set of tests were conducted to obtain an approximately optimal configuration for the choice of descriptive values used to rank the distributions and the subsequent effect on sampling efficiency, resulting in the use of mean redshift of a subset of tomographic bins, $\langle z \rangle_n$ being the choice of ranking parameter which gives the best efficiency (lowest  number of likelihood evaluations per posterior sample required for convergence of the chain).
As estimation of the minimum number of samples required for posterior estimates to become less noisy that the typical sampling noise in standard \deltaz marginalisation is also provided for the expected photometric redshift uncertainties of source distributions of the DES-Y3 analysis.

\response{Tests were conducted simulating a cosmic shear analysis where only a subset of cosmological and systematic parameters are inferred, compared to a full cosmic shear plus galaxy clustering case.
Despite this, \hyperrank is not limited to cosmic shear analysis and can be used without significant modifications on cosmic shear plus galaxy clustering (3x2pt) analysis.
We do not expect our conclusions to vary significantly for 3x2pt analysis.
Similarly, while tests here focus on the propagation of uncertainty from \emph{source} galaxy redshift distributions, \hyperrank can be used to simultaneously and independently propagate uncertainties from the \emph{lens} sample of galaxies for galaxy clustering plus tangential shear (2x2pt) and 3x2pt.}

For the particular levels of uncertainty expected for the DES-Y3 analysis we showed that the difference in obtained confidence contours between the standard approach using \deltaz shifts and \hyperrank are small and hence concluded that \deltaz was sufficient for the requirements of DES-Y3.
For the level of uncertainties present in DES-Y3 we have demonstrated that it is satisfactory to use the \deltaz approach which, whilst not as accurate as the \hyperrank approach, typically allows for faster convergence of the Monte Carlo inference chains, as can be seen as the dashed horizontal lines in \cref{fig:sampling_efficiency} which show the efficiencies for \deltaz.

\hyperrank provides a well-motivated approach for marginalising over the redshift distribution uncertainty affecting cosmological galaxy clustering and weak lensing surveys. It is nominally capable of marginalising over any potential form of such an uncertainty, subject to the ability to generate realisations samples of possible \nz using a model for the uncertainty. It thus also provides a much more complete and flexible approach to the commonly used and ad-hoc \deltaz approach, whilst still being able to contain that particular model and replicate findings made using it.

\section*{Acknowledgements}
This research manuscript made use of Astropy \citep{astropy:2013,astropy:2018}, ChainConsumer\footnote{samreay.github.io/ChainConsumer} \citep{Hinton16} and Matplotlib \citep{matplotlib}, and has been prepared using NASA's Astrophysics Data System Bibliographic Services.

IH, RR, SB acknowledge support from the European Research Council in the form of a Consolidator Grant with number 681431.

IH also acknowledges support from the Beecroft Trust.

JPC acknowledges support granted by Agencia Nacional de Investigaci\'on y Desarrollo (ANID) DOCTORADO BECAS CHILE/2016 - 72170279.

Funding for the DES Projects has been provided by the U.S. Department of Energy, the U.S. National Science Foundation, the Ministry of Science and Education of Spain, 
the Science and Technology Facilities Council of the United Kingdom, the Higher Education Funding Council for England, the National Center for Supercomputing 
Applications at the University of Illinois at Urbana-Champaign, the Kavli Institute of Cosmological Physics at the University of Chicago, 
the Center for Cosmology and Astro-Particle Physics at the Ohio State University,
the Mitchell Institute for Fundamental Physics and Astronomy at Texas A\&M University, Financiadora de Estudos e Projetos, 
Funda{\c c}{\~a}o Carlos Chagas Filho de Amparo {\`a} Pesquisa do Estado do Rio de Janeiro, Conselho Nacional de Desenvolvimento Cient{\'i}fico e Tecnol{\'o}gico and 
the Minist{\'e}rio da Ci{\^e}ncia, Tecnologia e Inova{\c c}{\~a}o, the Deutsche Forschungsgemeinschaft and the Collaborating Institutions in the Dark Energy Survey. 

The Collaborating Institutions are Argonne National Laboratory, the University of California at Santa Cruz, the University of Cambridge, Centro de Investigaciones Energ{\'e}ticas, 
Medioambientales y Tecnol{\'o}gicas-Madrid, the University of Chicago, University College London, the DES-Brazil Consortium, the University of Edinburgh, 
the Eidgen{\"o}ssische Technische Hochschule (ETH) Z{\"u}rich, 
Fermi National Accelerator Laboratory, the University of Illinois at Urbana-Champaign, the Institut de Ci{\`e}ncies de l'Espai (IEEC/CSIC), 
the Institut de F{\'i}sica d'Altes Energies, Lawrence Berkeley National Laboratory, the Ludwig-Maximilians Universit{\"a}t M{\"u}nchen and the associated Excellence Cluster Universe, 
the University of Michigan, NSF's NOIRLab, the University of Nottingham, The Ohio State University, the University of Pennsylvania, the University of Portsmouth, 
SLAC National Accelerator Laboratory, Stanford University, the University of Sussex, Texas A\&M University, and the OzDES Membership Consortium.

Based in part on observations at Cerro Tololo Inter-American Observatory at NSF's NOIRLab (NOIRLab Prop. ID 2012B-0001; PI: J. Frieman), which is managed by the Association of Universities for Research in Astronomy (AURA) under a cooperative agreement with the National Science Foundation.

The DES data management system is supported by the National Science Foundation under Grant Numbers AST-1138766 and AST-1536171.
The DES participants from Spanish institutions are partially supported by MICINN under grants ESP2017-89838, PGC2018-094773, PGC2018-102021, SEV-2016-0588, SEV-2016-0597, and MDM-2015-0509, some of which include ERDF funds from the European Union. IFAE is partially funded by the CERCA program of the Generalitat de Catalunya.
Research leading to these results has received funding from the European Research
Council under the European Union's Seventh Framework Program (FP7/2007-2013) including ERC grant agreements 240672, 291329, and 306478.
We  acknowledge support from the Brazilian Instituto Nacional de Ci\^encia
e Tecnologia (INCT) do e-Universo (CNPq grant 465376/2014-2).

This manuscript has been authored by Fermi Research Alliance, LLC under Contract No. DE-AC02-07CH11359 with the U.S. Department of Energy, Office of Science, Office of High Energy Physics.

\section*{Data Availability}
The DES Y3 data products used in this work, full
ensemble of DES Y3 source galaxy redshift distributions, chains and data products of the tests conducted here will be made publicly available following publication,
at the URL \url{https://des.ncsa.illinois.edu/releases}.

The \hyperrank code is available in the \cosmosis standard library repository 
\url{https://bitbucket.org/joezuntz/cosmosis-standard-library/src/des-y3/}.




\bibliographystyle{mn2e_2author_arxiv}
\bibliography{biblio,des_y3kp} 

\begin{thebibliography}{45}
\providecommand{\natexlab}[1]{#1}
\providecommand{\url}[1]{\texttt{#1}}
\providecommand{\urlprefix}{URL }
\providecommand{\eprint}[1][]{\url{#1}}

\bibitem[{{Abazajian} et~al.(2009){Abazajian}, {Adelman-McCarthy}
  et~al.}]{2009ApJS..182..543A}
{Abazajian}, K.~N., {Adelman-McCarthy}, J.~K., {Ag{\"u}eros}, M.~A., et~al.,
  2009, \apjs, 182, 2, 543, \eprint arXiv:{0812.0649}

\bibitem[{{Amon} et~al.(2021)}]{y3-cosmicshear1}
{Amon}, A., et~al., 2021, Submitted to PRD, \eprint arXiv:{2105.13543}

\bibitem[{{Astropy Collaboration} \& {Astropy
  Contributors}(2018)}]{astropy:2018}
{Astropy Collaboration}, 2018, \aj, 156, 3, 123, \eprint arXiv:{1801.02634}

\bibitem[{{Astropy Collaboration}(2013){Astropy Collaboration}, {Robitaille}
  et~al.}]{astropy:2013}
{Astropy Collaboration}, 2013, \aap, 558, A33, \eprint arXiv:{1307.6212}

\bibitem[{{Blanton} et~al.(2005){Blanton}, {Schlegel}
  et~al.}]{2005AJ....129.2562B}
{Blanton}, M.~R., {Schlegel}, D.~J., {Strauss}, M.~A., et~al., 2005, \aj, 129,
  6, 2562, \eprint arXiv:{astro-ph/0410166}

\bibitem[{{Bonnett} et~al.(2016){Bonnett}, {Troxel}
  et~al.}]{2016PhRvD..94d2005B}
{Bonnett}, C., {Troxel}, M.~A., {Hartley}, W., et~al., 2016, \prd, 94, 4,
  042005, \eprint arXiv:{1507.05909}

\bibitem[{Burkard \& Derigs(1980)}]{Burkard1980}
Burkard, R.~E., Derigs, U., 1980, The Linear Sum Assignment Problem, 1--15,
  Springer Berlin Heidelberg, Berlin, Heidelberg

\bibitem[{Cawthon et~al.(2020)}]{y3-lenswz}
Cawthon, R., et~al., 2020, Submitted to MNRAS, \eprint arXiv:{2012.12826}

\bibitem[{{Dark Energy Survey Collaboration}(2018){Dark Energy Survey
  Collaboration}, {Abbott} et~al.}]{2018PhRvD..98d3526A}
{Dark Energy Survey Collaboration}, 2018, \prd, 98, 4, 043526, \eprint
  arXiv:{1708.01530}

\bibitem[{{DeRose} et~al.(2019){DeRose}, {Wechsler}
  et~al.}]{2019arXiv190102401D}
{DeRose}, J., {Wechsler}, R.~H., {Becker}, M.~R., et~al., 2019, arXiv e-prints,
  \eprint arXiv:{1901.02401}

\bibitem[{{DeRose} et~al.(2021){DeRose}, {Wechsler} et~al.}]{y3-simvalidation}
{DeRose}, J., {Wechsler}, R.~H., {Becker}, M.~R., et~al., 2021, arXiv e-prints,
  \eprint arXiv:{2105.13547}

\bibitem[{{DES Collaboration}(2021){DES Collaboration}, {Abbott}
  et~al.}]{y3-3x2ptkp}
{DES Collaboration}, 2021, arXiv e-prints, \eprint arXiv:{2105.13549}

\bibitem[{Everett et~al.(2020)}]{y3-balrog}
Everett, S., et~al., 2020, Submitted to ApJS, \eprint arXiv:{2012.12825}

\bibitem[{{Feroz} et~al.(2009){Feroz}, {Hobson} \&
  {Bridges}}]{2009MNRAS.398.1601F}
{Feroz}, F., {Hobson}, M.~P., {Bridges}, M., 2009, \mnras, 398, 4, 1601,
  \eprint arXiv:{0809.3437}

\bibitem[{{Foreman-Mackey} et~al.(2013){Foreman-Mackey}, {Hogg}, {Lang} \&
  {Goodman}}]{2013PASP..125..306F}
{Foreman-Mackey}, D., {Hogg}, D.~W., {Lang}, D., {Goodman}, J., 2013, \pasp,
  125, 925, 306, \eprint arXiv:{1202.3665}

\bibitem[{Gatti et~al.(2020)Gatti, Giannini et~al.}]{y3-sourcewz}
Gatti, M., Giannini, G., et~al., 2020, Submitted to MNRAS, \eprint
  arXiv:{2012.08569}

\bibitem[{{Hadzhiyska} et~al.(2020){Hadzhiyska}, {Alonso}, {Nicola} \&
  {Slosar}}]{2020JCAP...10..056H}
{Hadzhiyska}, B., {Alonso}, D., {Nicola}, A., {Slosar}, A., 2020, \jcap, 2020,
  10, 056, \eprint arXiv:{2007.14989}

\bibitem[{{Hamana} et~al.(2020){Hamana}, {Shirasaki}
  et~al.}]{2020PASJ...72...16H}
{Hamana}, T., {Shirasaki}, M., {Miyazaki}, S., et~al., 2020, \pasj, 72, 1, 16,
  \eprint arXiv:{1906.06041}

\bibitem[{{Handley} et~al.(2015){Handley}, {Hobson} \&
  {Lasenby}}]{2015MNRAS.450L..61H}
{Handley}, W.~J., {Hobson}, M.~P., {Lasenby}, A.~N., 2015, \mnras, 450, L61,
  \eprint arXiv:{1502.01856}

\bibitem[{{Hartley} et~al.(2020){Hartley}, {Chang}
  et~al.}]{2020MNRAS.496.4769H}
{Hartley}, W.~G., {Chang}, C., {Samani}, S., et~al., 2020, \mnras, 496, 4,
  4769, \eprint arXiv:{2003.10454}

\bibitem[{Hartley et~al.(2020)Hartley, Choi et~al.}]{y3-deepfields}
Hartley, W.~G., Choi, A., et~al., 2020, Submitted to MNRAS, \eprint
  arXiv:{2012.12824}

\bibitem[{{Heymans} et~al.(2021){Heymans}, {Tr{\"o}ster}
  et~al.}]{2020arXiv200715632H}
{Heymans}, C., {Tr{\"o}ster}, T., {Asgari}, M., et~al., 2021, \aap, 646, A140,
  \eprint arXiv:{2007.15632}

\bibitem[{{Hikage} et~al.(2019){Hikage}, {Oguri} et~al.}]{2019PASJ...71...43H}
{Hikage}, C., {Oguri}, M., {Hamana}, T., et~al., 2019, \pasj, 71, 2, 43,
  \eprint arXiv:{1809.09148}

\bibitem[{{Hildebrandt} et~al.(2017){Hildebrandt}, {Viola}
  et~al.}]{2017MNRAS.465.1454H}
{Hildebrandt}, H., {Viola}, M., {Heymans}, C., et~al., 2017, \mnras, 465, 2,
  1454, \eprint arXiv:{1606.05338}

\bibitem[{{Hinton}(2016)}]{Hinton16}
{Hinton}, S.~R., 2016, The Journal of Open Source Software, 1, 4, 00045

\bibitem[{{Hoyle} et~al.(2018){Hoyle}, {Gruen} et~al.}]{2018MNRAS.478..592H}
{Hoyle}, B., {Gruen}, D., {Bernstein}, G.~M., et~al., 2018, \mnras, 478, 1,
  592, \eprint arXiv:{1708.01532}

\bibitem[{{Hu}(1999)}]{1999ApJ...522L..21H}
{Hu}, W., 1999, \apjl, 522, 1, L21, \eprint arXiv:{astro-ph/9904153}

\bibitem[{Hunter(2007)}]{matplotlib}
Hunter, J.~D., 2007, Computing in Science \& Engineering, 9, 3, 90

\bibitem[{{Joachimi} et~al.(2021){Joachimi}, {Lin}
  et~al.}]{2020arXiv200701844J}
{Joachimi}, B., {Lin}, C.~A., {Asgari}, M., et~al., 2021, \aap, 646, A129,
  \eprint arXiv:{2007.01844}

\bibitem[{{Joudaki} et~al.(2020){Joudaki}, {Hildebrandt}
  et~al.}]{2020A&A...638L...1J}
{Joudaki}, S., {Hildebrandt}, H., {Traykova}, D., et~al., 2020, \aap, 638, L1,
  \eprint arXiv:{1906.09262}

\bibitem[{{Kitching} \& {Taylor}(2011)}]{2011MNRAS.410.1677K}
{Kitching}, T.~D., {Taylor}, A.~N., 2011, \mnras, 410, 3, 1677, \eprint
  arXiv:{1005.2063}

\bibitem[{{Leistedt} et~al.(2016){Leistedt}, {Mortlock} \&
  {Peiris}}]{2016MNRAS.460.4258L}
{Leistedt}, B., {Mortlock}, D.~J., {Peiris}, H.~V., 2016, \mnras, 460, 4, 4258,
  \eprint arXiv:{1602.05960}

\bibitem[{{Masters} et~al.(2015){Masters}, {Capak}
  et~al.}]{2015ApJ...813...53M}
{Masters}, D., {Capak}, P., {Stern}, D., et~al., 2015, \apj, 813, 1, 53,
  \eprint arXiv:{1509.03318}

\bibitem[{Myles et~al.(2020)Myles, Alarcon et~al.}]{y3-sompz}
Myles, J., Alarcon, A., et~al., 2020, Submitted to MNRAS, \eprint
  arXiv:{2012.08566}

\bibitem[{{Rau} et~al.(2020){Rau}, {Wilson} \&
  {Mandelbaum}}]{2020MNRAS.491.4768R}
{Rau}, M.~M., {Wilson}, S., {Mandelbaum}, R., 2020, \mnras, 491, 4, 4768,
  \eprint arXiv:{1904.09988}

\bibitem[{{S{\'a}nchez} \& {Bernstein}(2019)}]{2019MNRAS.483.2801S}
{S{\'a}nchez}, C., {Bernstein}, G.~M., 2019, \mnras, 483, 2, 2801, \eprint
  arXiv:{1807.11873}

\bibitem[{{S{\'a}nchez} et~al.(2020){S{\'a}nchez}, {Raveri}, {Alarcon} \&
  {Bernstein}}]{2020MNRAS.498.2984S}
{S{\'a}nchez}, C., {Raveri}, M., {Alarcon}, A., {Bernstein}, G.~M., 2020,
  \mnras, 498, 2, 2984, \eprint arXiv:{2004.09542}

\bibitem[{{Schmidt} et~al.(2020){Schmidt}, {Malz} et~al.}]{2020MNRAS.499.1587S}
{Schmidt}, S.~J., {Malz}, A.~I., {Soo}, J.~Y.~H., et~al., 2020, \mnras, 499, 2,
  1587, \eprint arXiv:{2001.03621}

\bibitem[{{Secco} et~al.(2021){Secco}, {Samuroff} et~al.}]{y3-cosmicshear2}
{Secco}, L.~F., {Samuroff}, S., {Krause}, E., et~al., 2021, arXiv e-prints,
  \eprint arXiv:{2105.13544}

\bibitem[{{Springel}(2005)}]{2005MNRAS.364.1105S}
{Springel}, V., 2005, \mnras, 364, 4, 1105, \eprint arXiv:{astro-ph/0505010}

\bibitem[{{St{\"o}lzner} et~al.(2021){St{\"o}lzner}, {Joachimi}, {Korn},
  {Hildebrandt} \& {Wright}}]{2020arXiv201207707S}
{St{\"o}lzner}, B., {Joachimi}, B., {Korn}, A., {Hildebrandt}, H., {Wright},
  A.~H., 2021, \aap, 650, A148, \eprint arXiv:{2012.07707}

\bibitem[{{Taylor} \& {Kitching}(2010)}]{2010MNRAS.408..865T}
{Taylor}, A.~N., {Kitching}, T.~D., 2010, \mnras, 408, 2, 865, \eprint
  arXiv:{1003.1136}

\bibitem[{{Tessore} \& {Harrison}(2020)}]{2020OJAp....3E...6T}
{Tessore}, N., {Harrison}, I., 2020, The Open Journal of Astrophysics, 3, 1, 6,
  \eprint arXiv:{2003.11558}

\bibitem[{{Wright} et~al.(2020){Wright}, {Hildebrandt}
  et~al.}]{2020A&A...640L..14W}
{Wright}, A.~H., {Hildebrandt}, H., {van den Busch}, J.~L., et~al., 2020, \aap,
  640, L14, \eprint arXiv:{2005.04207}

\bibitem[{{Zuntz} et~al.(2015){Zuntz}, {Paterno} et~al.}]{2015A&C....12...45Z}
{Zuntz}, J., {Paterno}, M., {Jennings}, E., et~al., 2015, Astronomy and
  Computing, 12, 45, \eprint arXiv:{1409.3409}

\end{thebibliography}




\section*{Affiliations}
\label{section:affiliations}

$^{1}$ Jodrell Bank Centre for Astrophysics, Department of Physics and Astronomy, The University of Manchester, Manchester M13 9PL, UK\\
$^{2}$ Department of Physics, University of Oxford, Denys Wilkinson Building, Keble Road, Oxford OX1 3RH, UK\\
$^{3}$ Institute for Astronomy, University of Edinburgh, Edinburgh EH9 3HJ, UK\\
$^{4}$ Department of Physics and Astronomy, University of Pennsylvania, Philadelphia, PA 19104, USA\\
$^{5}$ Argonne National Laboratory, 9700 South Cass Avenue, Lemont, IL 60439, USA\\
$^{6}$ Institute of Space Sciences (ICE, CSIC), Campus UAB, Carrer de Can Magrans, s/n, 08193 Barcelona, Spain \\
$^{7}$ Institut d'Estudis Espacials de Catalunya (IEEC), E-08034 Barcelona, Spain \\
$^{8}$ Department of Physics, University of Michigan, Ann Arbor, MI 48109, USA\\
$^{9}$ Instituto de F\'{i}sica Te\'orica, Universidade Estadual Paulista, S\~ao Paulo, Brazil\\
$^{10}$ Department of Physics, Stanford University, 382 Via Pueblo Mall, Stanford, CA 94305, USA\\
$^{11}$ Kavli Institute for Particle Astrophysics \& Cosmology, P. O. Box 2450, Stanford University, Stanford, CA 94305, USA\\
$^{12}$ SLAC National Accelerator Laboratory, Menlo Park, CA 94025, USA\\
$^{13}$ Laborat\'orio Interinstitucional de e-Astronomia - LIneA, Rua Gal. Jos\'e Cristino 77, Rio de Janeiro, RJ - 20921-400, Brazil\\
$^{14}$ Department of Physics, Carnegie Mellon University, Pittsburgh, Pennsylvania 15312, USA\\
$^{15}$ Center for Cosmology and Astro-Particle Physics, The Ohio State University, Columbus, OH 43210, USA\\
$^{16}$ Santa Cruz Institute for Particle Physics, Santa Cruz, CA 95064, USA\\
$^{17}$ Department of Astronomy, University of California, Berkeley,  501 Campbell Hall, Berkeley, CA 94720, USA\\
$^{18}$ Jet Propulsion Laboratory, California Institute of Technology, 4800 Oak Grove Dr., Pasadena, CA 91109, USA\\
$^{19}$ Department of Astronomy/Steward Observatory, University of Arizona, 933 North Cherry Avenue, Tucson, AZ 85721-0065, USA\\
$^{20}$ Kavli Institute for Cosmology, University of Cambridge, Madingley Road, Cambridge CB3 0HA, UK\\
$^{21}$ Churchill College, University of Cambridge, CB3 0DS Cambridge, UK\\
$^{22}$ Department of Astronomy, University of Illinois at Urbana-Champaign, 1002 W. Green Street, Urbana, IL 61801, USA\\
$^{23}$ Center for Astrophysical Surveys, National Center for Supercomputing Applications, 1205 West Clark St., Urbana, IL 61801, USA\\
$^{24}$ D\'{e}partement de Physique Th\'{e}orique and Center for Astroparticle Physics, Universit\'{e} de Gen\`{e}ve, 24 quai Ernest Ansermet, CH-1211 Geneva, Switzerland\\
$^{25}$ Fermi National Accelerator Laboratory, P. O. Box 500, Batavia, IL 60510, USA\\
$^{26}$ Department of Applied Mathematics and Theoretical Physics, University of Cambridge, Cambridge CB3 0WA, UK\\
$^{27}$ Kavli Institute for Cosmological Physics, University of Chicago, Chicago, IL 60637, USA\\
$^{28}$ ICTP South American Institute for Fundamental Research\\ Instituto de F\'{\i}sica Te\'orica, Universidade Estadual Paulista, S\~ao Paulo, Brazil\\
$^{29}$ Centro de Investigaciones Energ\'eticas, Medioambientales y Tecnol\'ogicas (CIEMAT), Madrid, Spain\\
$^{30}$ Brookhaven National Laboratory, Bldg 510, Upton, NY 11973, USA\\
$^{31}$ Department of Physics, Duke University Durham, NC 27708, USA\\
$^{32}$ Departamento de F\'isica Matem\'atica, Instituto de F\'isica, Universidade de S\~ao Paulo, CP 66318, S\~ao Paulo, SP, 05314-970, Brazil\\
$^{33}$ CNRS, UMR 7095, Institut d'Astrophysique de Paris, F-75014, Paris, France\\
$^{34}$ Sorbonne Universit\'es, UPMC Univ Paris 06, UMR 7095, Institut d'Astrophysique de Paris, F-75014, Paris, France\\
$^{35}$ Department of Physics \& Astronomy, University College London, Gower Street, London, WC1E 6BT, UK\\
$^{36}$ Instituto de Astrofisica de Canarias, E-38205 La Laguna, Tenerife, Spain\\
$^{37}$ Universidad de La Laguna, Dpto. Astrof\'isica, E-38206 La Laguna, Tenerife, Spain\\
$^{38}$ Institut de F\'{\i}sica d'Altes Energies (IFAE), The Barcelona Institute of Science and Technology, Campus UAB, 08193 Bellaterra (Barcelona) Spain\\
$^{39}$ Physics Department, 2320 Chamberlin Hall, University of Wisconsin-Madison, 1150 University Avenue Madison, WI  53706-1390\\
$^{40}$ INAF-Osservatorio Astronomico di Trieste, via G. B. Tiepolo 11, I-34143 Trieste, Italy\\
$^{41}$ Institute for Fundamental Physics of the Universe, Via Beirut 2, 34014 Trieste, Italy\\
$^{42}$ Faculty of Physics, Ludwig-Maximilians-Universit\"at, Scheinerstr. 1, 81679 Munich, Germany\\
$^{43}$ Department of Physics, The Ohio State University, Columbus, OH 43210, USA\\
$^{44}$ Institute of Theoretical Astrophysics, University of Oslo. P.O. Box 1029 Blindern, NO-0315 Oslo, Norway\\
$^{45}$ Instituto de Fisica Teorica UAM/CSIC, Universidad Autonoma de Madrid, 28049 Madrid, Spain\\
$^{46}$ Department of Astronomy, University of Michigan, Ann Arbor, MI 48109, USA\\
$^{47}$ School of Mathematics and Physics, University of Queensland,  Brisbane, QLD 4072, Australia\\
$^{48}$ Max Planck Institute for Extraterrestrial Physics, Giessenbachstrasse, 85748 Garching, Germany\\
$^{49}$ Universit\"ats-Sternwarte, Fakult\"at f\"ur Physik, Ludwig-Maximilians Universit\"at M\"unchen, Scheinerstr. 1, 81679 M\"unchen, Germany\\
$^{50}$ Center for Astrophysics $\vert$ Harvard \& Smithsonian, 60 Garden Street, Cambridge, MA 02138, USA\\
$^{51}$ Australian Astronomical Optics, Macquarie University, North Ryde, NSW 2113, Australia\\
$^{52}$ Lowell Observatory, 1400 Mars Hill Rd, Flagstaff, AZ 86001, USA\\
$^{53}$ Instituci\'o Catalana de Recerca i Estudis Avançats, E-08010 Barcelona, Spain\\
$^{54}$ Perimeter Institute for Theoretical Physics, 31 Caroline St. North, Waterloo, ON N2L 2Y5, Canada\\
$^{55}$ Institute of Astronomy, University of Cambridge, Madingley Road, Cambridge CB3 0HA, UK\\
$^{56}$ Observat\'orio Nacional, Rua Gal. Jos\'e Cristino 77, Rio de Janeiro, RJ - 20921-400, Brazil\\
$^{57}$ Department of Astrophysical Sciences, Princeton University, Peyton Hall, Princeton, NJ 08544, USA\\
$^{58}$ School of Physics and Astronomy, University of Southampton,  Southampton, SO17 1BJ, UK\\
$^{59}$ Computer Science and Mathematics Division, Oak Ridge National Laboratory, Oak Ridge, TN 37831\\
$^{60}$ Institute of Cosmology and Gravitation, University of Portsmouth, Portsmouth, PO1 3FX, UK



\appendix



\bsp	
\label{lastpage}
\end{document}